\documentclass[8pt]{article}
\RequirePackage[OT1]{fontenc}
\usepackage{natbib}
\RequirePackage[colorlinks,citecolor=blue,urlcolor=blue]{hyperref}
\usepackage[english]{babel}
\usepackage{amsthm}
\usepackage{amsmath}
\usepackage{amsfonts}
\usepackage{amssymb}
\usepackage{graphicx}
\usepackage{enumerate}
\usepackage{color}
\usepackage{tikz}
\usepackage{array}
\usepackage{multirow}

\def\frac#1#2{{\textstyle{#1\over#2}}}

\DeclareSymbolFont{AMSb}{U}{msb}{m}{n}
\DeclareMathSymbol{\Natural}{\mathbin}{AMSb}{"4E}
\DeclareMathSymbol{\Integer}{\mathbin}{AMSb}{"5A}
\DeclareMathSymbol{\Real}{\mathbin}{AMSb}{"52}
\DeclareMathSymbol{\Rational}{\mathbin}{AMSb}{"51}
\DeclareMathSymbol{\Imaginary}{\mathbin}{AMSb}{"49}
\DeclareMathSymbol{\Complex}{\mathbin}{AMSb}{"43} 
\DeclareMathSymbol{\Disk}{\mathbin}{AMSb}{"44} 
\def\bi{\begin{itemize}}
\def\ei{\end{itemize}}
\def\bd{\begin{description}}
\def\ed{\end{description}}
\def\ben{\begin{enumerate}}
\def\een{\end{enumerate}}
\def\bv{\begin{verbatim}}
\def\ev{\end{verbatim}}

\def\hat#1{{\widehat{#1}}}

\newcommand{\bs}{\boldsymbol}

\def\pr{{\rm Pr}}
\def\Pr{\pr}
\def\E{{\rm E}}

\def\var{{\rm var}}

\def\cov{{\rm cov}}

\def\2to{{\ {\buildrel 2\over \longrightarrow}\ }}

\def\I1ton{{$I_1,\ldots,I_n$}}
\def\X1ton{{$X_1,\ldots,X_n$}}
\def\Y1ton{{$Y_1,\ldots,Y_n$}}
\def\Z1ton{{$Z_1,\ldots,Z_n$}}
\def\R1ton{{$R_1,\ldots,R_n$}}
\def\e1ton{{$e_1,\ldots,e_n$}}
\def\t1ton{{$t_1,\ldots,t_n$}}
\def\x1ton{{$x_1,\ldots,x_n$}}
\def\y1ton{{$y_1,\ldots,y_n$}}
\def\z1ton{{$z_1,\ldots,z_n$}}

\allowdisplaybreaks
\usepackage{float}
\usepackage[algoruled,boxed,lined]{algorithm2e} 
\makeatletter 
\g@addto@macro{\@algocf@init}{\SetKwInOut{Parameter}{Parameters}} 
\makeatother
 \usepackage{mathtools}
 
 \usepackage{verbatim}

\definecolor{myorange}{HTML}{E24100}
 
\begin{document}
\thispagestyle{empty}
\baselineskip=28pt
\vskip 5mm
\begin{center} {\Large{\bf Flexible Modeling of Multivariate Spatial Extremes}}
\end{center}

\baselineskip=12pt
\vskip 5mm
\begin{center}
\large
Yan Gong$^1$ and Rapha\"el Huser$^1$
\end{center}
\footnotetext[1]{
\baselineskip=10pt Statistics Program, Computer, Electrical and Mathematical Sciences and Engineering (CEMSE) Division, King Abdullah University of Science and Technology (KAUST), Thuwal 23955-6900, Saudi Arabia. E-mail: yan.gong@kaust.edu.sa, raphael.huser@kaust.edu.sa.}
\baselineskip=17pt
\vskip 4mm
\centerline{\today}
\vskip 6mm
\begin{center}
{\large{\bf Abstract}}
\end{center}


We {develop} a novel multi-factor copula model for multivariate spatial extremes, which is designed to capture the different combinations of marginal and cross-extremal dependence structures within and across different spatial random fields. Our {proposed model, which can be seen as a multi-factor copula model,} can capture all possible distinct combinations of extremal dependence structures within each individual spatial process while allowing flexible cross-process extremal dependence structures for both upper and lower tails. We {show how to perform} Bayesian inference {for the proposed model} using a Markov chain Monte Carlo algorithm based on carefully designed block proposals with an adaptive step size. In our real data application, we apply our model to study the upper and lower extremal dependence structures of the daily maximum air temperature (TMAX) and daily minimum air temperature (TMIN) from the state of Alabama in the southeastern United States. {The fitted multivariate spatial model is found to provide a good fit in the lower and upper joint tails, both in terms of the spatial dependence structure within each individual process, as well as in terms of the cross-process dependence structure. Our results suggest that the TMAX and TMIN processes are quite strongly spatially dependent over the state of Alabama, and moderately cross-dependent. From a practical perspective, this implies that it may be worthwhile to model them jointly when interest lies in a computing spatial risk measures that involve both quantities.}

\baselineskip=16pt
\par\vfill\noindent
\pagenumbering{arabic}
\baselineskip=24pt
\newpage

                                                    \section{Introduction}\label{sec:Introduction}
The statistical modeling of multivariate spatial extremes is a fundamental but challenging task in geophysical, ecological, environmental, and climate sciences, in order to quantify the probability and compound risks of natural hazards such as floods, droughts, heat waves{, and strong winds. In a related context, model-based statistical approaches can also help understand and disentangle} the probabilistic behavior of anthropogenic processes, such as air pollutant concentrations and their impact on the environment. {In practice, multiple} variables {are often} observed simultaneously from monitors or generated by climate or geophysical models at different locations{, and it is often the co-occurrence of extremes from these multiple variables that lead to the biggest impacts \citep{zscheischler2020typology, raymond2020understanding}. For example, high temperatures combined with strong wind might increase wild fire risk \citep{littell2016review, ruffault2018extreme}; the heat index, which may be used as a measure of climate ``livability'', is large when both temperature and relative humidity are simultaneously high; similarly, a heat wave is defined by the Australian Bureau of Meteorology as a multivariate event where ``the maximum and the minimum temperatures are unusually hot over a three-day period at a location'' (\href{http://www.bom.gov.au/australia/heat wave/knowledge-centre/understanding.shtml#:~:text=What%20is%20a%20heat wave%3F,past%20weather%20at%20the%20location.}{link})
, see also \citet{baldwin2019temporally} and \citet{perkins2020increasing} for alternative heat wave definitions; and the formation of a hurricane is known to involve both warm ocean water, as well as moist, humid air in the same spatial region.} Thus, it has become increasingly important to model such complex multivariate spatial extreme data.

In the literature of multivariate spatial extremes modeling, recent work has been done {to extend} spatial max-stable processes to the multivariate setting. In particular, \cite{genton2015multivariate} proposed several multivariate {spatial max-stable} processes{, generalizing} the theory and application of max-stable processes to the multivariate context. Specifically, they introduced multivariate versions of (i) the Smith Gaussian extreme-value model \citep{smith1990max}, (ii) the Schlather extremal-Gaussian model \citep{schlather2002models} {and its} extremal-$t$ model {extension} \citep{opitz2013extremal}, and (iii) the Brown--Resnick model \citep{brown1977extreme,kabluchko2009stationary}. {Later, \cite{oesting2017statistical} introduced a bivariate Brown--Resnick process, constructed from a customized pseudo cross-variogram, as a joint spatial model for real extreme observations and model-based forecasts, and designed to post-process and enhance probabilistic weather extremes forecasts}. More recently, \cite{vettori2019bayesian2} developed nested tree-based multivariate max-stable processes, to study extreme concentrations of multiple pollutants over a spatial domain. The construction of {these} nested max-stable processes extends the spatial Reich--Shaby process \citep{reich2012hierarchical}, which has a conditional independence representation given some latent positive stable random effects and is jointly max-stable (both across space and variables). Max-stable processes have been used in the above studies on the basis that they have a strong theoretical motivation and {well-understood} tail properties; however, {max-stable processes are only justified for block maxima, and their asymptotic characterization entails a fairly rigid tail structure; see, e.g., \citet{huser2022advances} for a detailed review. In particular, these processes are always asymptotically dependent (AD), which can be a strong limitation in some applications}. This {property} means that if $Z(\bs s)$ is a process with common margins and infinite upper endpoint, then $\chi=\lim_{z\to\infty}\pr\{Z(\bs s_1)>z\mid Z(\bs s_2)>z\}>0$ for any pair of sites $\{\bs s_1,\bs s_2\}$. However, empirical evidence suggests that {asymptotic independence} (AI) with $\chi=0$ is equally important in practice, particularly over large domains or with data exhibiting short-range extremal dependence through localized extreme events (e.g., convective precipitation events{, or wind gusts}); see, e.g., {\cite{wadsworth2012dependence}, \citet{Opitz:2016}, \citet{Hazra.etal:2021}, \citet{huser2022advances}, and \citet{zhang2022modeling}.} Therefore, from a methodological and practical perspective, it is crucial to develop flexible multivariate spatial models that can {capture} AD or AI in their upper and/or lower tails, with a smooth transition in between{, in order to model the full within- and cross-process dependence structures, while accurately capturing the joint behavior of extremes.}

In the classical geostatistical framework, several {types} of multivariate spatial models have been proposed, including multivariate Gaussian processes \citep{kleiber2012nonstationary,genton2015cross,cressie2016multivariate,gelfand2021multivariate} or multivariate {$t$-processes} \citep{kotz2004multivariate,opitz2013extremal, hazra2020multivariate}. However, while the former are always AI, the latter are always AD {for all pairs of sites}; thus, both lack tail flexibility. In addition, both Gaussian and $t$-processes are tail-symmetric, and there has been little work on constructing multivariate spatial models with flexible asymmetric tail dependencies.

In a recent related work, \cite{krupskii2019copula} proposed an exponential factor copula model for non-Gaussian multivariate spatial data{, which allows fast likelihood-based inference}. In their model, the obtained extremal dependence structure within or across fields can capture tail asymmetry, but it always exhibits AD, while AI lies on the boundary of the parameter space, which is a crucial practical limitation.


To capture complex extremal dependence structure in multivariate spatial extremes, we propose in this paper a novel multi-factor copula model for multivariate spatial data that can capture all possible distinct combinations of extremal dependence types within each individual spatial process while allowing for flexible cross-process extremal dependence structures, for both the upper and lower tails. Our new model builds upon \cite{huser2019modeling}, who proposed a spatial extremes model for the upper tail of a single variable, and it extends the bivariate copula model of \cite{gong2022asymmetric}{, used to estimate time-varying tail dependencies in bivariate financial time series,} to the multivariate spatial extremes setting. {Rewriting our proposed model as a Bayesian hierarchical model with multiple latent variables, we here perform Bayesian inference} using a customized Markov chain Monte Carlo (MCMC) algorithm based on carefully designed block proposals with an adaptive step size. Specially, we implement adaptive Metropolis--Hastings updates for hyperparameters and {use} the Metropolis-adjusted Langevin algorithm (MALA) to update the high-dimensional vector of latent variables.

We apply our model to study the lower and upper extremal dependence structures of the daily maximum air temperature (TMAX) and daily minimum air temperature (TMIN) in {the state of} Alabama in the southeastern United States (US). {Since} simultaneous temperature extremes (both high and low extremes) can severely affect human health \citep{vicedo2021burden, mitchell2021climate}, {the environment, ecosystems,} and energy consumption \citep{seneviratne2012changes}{, it is crucial to study the} co-occurence {probability of such extreme events} over space{, and we here exploit our model to assess the risk of jointly low/high daily temperatures in Alabama, a state known for its subtropical climate (due to its proximity to the Gulf of Mexico), which is prone to thunderstorms, tropical storms, and hurricanes}. 
{In other words, we use our proposed methodology to jointly model} the spatial extremal {dependence} within and across TMAX and TMIN random fields over the {whole} state of Alabama, US.

This paper is organized as follows. In Section \S \ref{sec:model}, we present the model construction and derive the model's extremal dependence properties. In Section \S \ref{sec:inference}, we describe the proposed Bayesian inference procedure, which is carefully designed for our model's {hierarchical} specification. A simulation study is discussed in Section \S \ref{sec:simulation} and the results of the real-data application are presented in Section \S \ref{sec:application}. In  Section \S \ref{sec:conclusion}, we discuss our results and findings, and conclude with some {perspectives} on future research.

                                         \section{Modeling}\label{sec:model}
In this section, we describe the construction of our proposed multivariate spatial model with flexible extremal dependence structure in both tails. The dependence (i.e., copula) model is designed to capture the different combinations of marginal and cross-extremal dependence structures within and across different spatial random fields. We present the construction steps in detail and derive the extremal dependence properties. Furthermore, we demonstrate the flexibility of the developed model in a bivariate spatial setting.

\subsection{Model construction}\label{sec:modelconstruction}
We define the $p$-dimensional multivariate random process $\bs X(\bs s)  = (X_1(\bs s),\ldots,X_p(\bs s))^\top$ on the spatial domain $\mathcal{S} \subset \Real^D$, with finite-dimensional realizations at the $d$ spatial locations $\{\bs s_1, \dots, \bs s_d\}\subset \mathcal{S}$, as
\begin{align}
X_i(\bs s) = \alpha_i\big\{ \gamma_i[\delta^UR_0^U+(1-\delta^U)R_i^U]-(1-\gamma_i)[\delta^LR_0^L+(1-\delta^L)R_i^L]\big\}+W_i(\bs s), \quad {i=1,\ldots,p,}
\label{equ:model_i}
\end{align}
where $\alpha_i>0$, and $\gamma_i, \delta^U, \delta^L \in [0,1]$ are dependence parameters. For simplicity, we write $X_{ij} := X_i(\bs s_j)$ and $\bs X_{i} := (X_{i1},\dots, X_{id})^\top$, and similarly for $W_{ij}$ and $\bs W_i$, $i=1, \ldots p$, $j=1,\ldots, d$. {The (spatially-constant) random variables} $R^U_{0}, R^L_{0}, \{R^U_{i}\}, \{R^L_{i}\} \stackrel{i.i.d.}{\sim} \rm Exp(1)$ are mutually independent {and exponentially distributed, while} $W(\bs s) = (W_1(\bs s), \ldots, W_p(\bs s))^\top$ is a multivariate random field displaying AI in both tails, with exponentially-decaying tails, and defined more precisely below. Note that although the {latent} random factors $\{R_i^U\}$ and $\{R_i^L\}$ are field-specific, $R_0^U$ and $R_0^L$ are shared among the $p$ different random fields, which induces cross-dependence (and potentially cross-extremal dependence). {Our proposed model \eqref{equ:model_i} is carefully parametrized, so that the} dependence parameters play different roles and control the {model's flexibility} in different ways, {thus ensuring that it remains theoretically identifiable}. In particular, $\alpha_i$ controls the overall “weight” of the spatially-constant common factors with respect to the non-trivial spatially correlated random field $W_i(\bs s)${, thus mainly controlling the ``overall'' spatial dependence strength of the $i$-th field while also impacting its extremal dependence type}; $\gamma_i$ captures the extent of tail asymmetry for  the $i$-th field{, with values $\gamma_i>0.5$ producing ``right-skewed'' dependencies with larger mass in the joint upper tail than in the joint lower tail, and vice versa when $\gamma_i<0.5$}; and $\delta^U/\delta^L$ (which are {here} common to all fields) control the strength of the upper/lower tail cross-dependence{, as $\delta^U$ (respectively $\delta^L$) determines the balance between the field-specific random effect $R_i^U$ (respectively $R_i^L$) and the shared random effect $R_0^U$ (respectively, $R_0^L$)}. 
We demonstrate the model's flexibility using simulation examples in \S \ref{subsec:illustration}.

Here, for simplicity, we assume that the processes $W_i(\bs s)$, $i=1,\ldots,p$,  are cross-correlated trans-Gaussian processes with standard Laplace $(0, 1)$ margins, i.e., $W_{ij}=W_i(\bs s_j)\sim F_W$, where 
{$F_{W}(w)=[1+\text{sign}(w)\{1-\exp(-|w|)\}]/2$, $w\in\mathbb{R}$} and the variables $W_{ij}$ are driven by a Gaussian copula. The cross-dependence between the process $W_i(\bs s)$ {may} be constructed using the {Gaussian} Linear Model of Coregionalization (LMC) \citep{bourgault1991multivariable}{, although there are also other possibilities}. Specifically, we first define the Gaussian random vector
$$
(\bs W'^\top_1,\dots, \bs W'^\top_p)^\top = L(\bs W_1^{*T},\dots, \bs W_p^{*T})^\top,
$$ where {$\bs W^*_i \sim \mathcal{N}_d (\bs 0, \Sigma_i^*)$ are independent $d$-dimensional Gaussian vectors with mean zero, variance one, and spatial correlation structure $\Sigma_i^*$;} the matrix $L$ is a lower triangular $p\times p$ matrix such that $\Sigma = LL^\top$ is the cross-covariance matrix of $(\bs W'^\top_1,\dots, \bs W'^\top_p)^\top$. Finally, we obtain $\bs W_i = (W_{i1}, \ldots, W_{id})^\top$ by back-transforming the correlated Gaussian vector $\bs W'_i$ to the Laplace scale, i.e.,
$$\bs W_i = F_W^{-1}\{\Phi(\bs W'_i)\}.$$
More details about the properties of $(\bs W'^\top_1,\dots, \bs W'^\top_p)^\top$ are given in Appendix \S\ref{appendix:sigma}. 

Our proposed dependence model is defined by extracting the copula from the multivariate spatial process (\ref{equ:model_i}). We denote the joint cumulative distribution function of the multivariate spatial ($p\times d$)-dimensional random vector $\bs X=(\bs X_1^\top,\dots,\bs X_p^\top)^\top$ as $\bs X\sim F^{\bs X}$. In addition, we define the {marginally uniform} random vector $\bs U=(\bs U_1,\dots,\bs U_p)^\top$, {with} $\bs U_i = (U_{i1}, \ldots U_{id})^\top$ {and} $U_{ij}=F^X_{i}(X_{ij})\sim{\rm Unif}(0,1)$ for $i=1,\dots,p$, $j=1,\dots,d$, where $F^X_i$ is the marginal cumulative distribution function of $\bs X_i$, which can be obtained in the closed form (see Appendix \S\ref{appendix:Fxi} for the precise mathematical expressions). Thus, the associated copula of $\bs X$ {(i.e., the joint distribution function of the vector $\bs U$)} and its density can be obtained {respectively} as
\begin{align}
\notag C^{\bs X}(\bs u_1,\dots,\bs u_p)=F^{\bs X}\{(F^X_{1})^{-1}(\bs u_1),\dots, (F^X_{p})^{-1}(\bs u_p)\},\\
c^{\bs X}(\bs u_1,\dots,\bs u_p)={f^{\bs X}\{(F^X_{1})^{-1}(\bs u_1),\dots, (F^X_{p})^{-1}(\bs u_p)\}\over f^X_{1}\{(F^X_{1})^{-1}(\bs u_1)\}\times\cdots\times f^X_{p}\{(F^X_{p})^{-1}(\bs u_p)\} },
\label{eq:CU4}
\end{align}
where $f_i^{X}$ and $f^{\bs X}$ are the marginal and joint densities of $\bs X_i$ and $\bs X$, respectively, and the functions are applied componentwise. Note that the quantile function $(F^X_{i})^{-1}$ is not available in closed form, but it can be approximated {relatively quickly and} accurately using {standard} numerical root-finding algorithms.

\subsection{Extremal dependence properties}\label{sec:extremaldep}
The proposed model is designed to capture different combinations of marginal and cross-extremal dependence structures within and {across} the different random fields. In particular, the model can capture all the possible distinct combinations of extremal dependence types marginally, while allowing high flexibility in capturing cross-process extremal dependence for both upper and lower tails.

The model can be viewed as a particular type of random scale construction  \citep{engelke2019extremal, huser2019modeling}, and the marginal and cross-extremal {dependencies of} model (\ref{equ:model_i}) {may be classified} as reported in Table \ref{tab:tailproperties}. See Appendix \S\ref{appendix:proof} for the proof. {In the bivariate spatial setting (i.e., with $p=2$),} there are a number of $2\times2\times2 = 8$ different combinations of all possible marginal and cross-extremal dependence scenarios for each tail (AI or AD for the first process $\times$ AI or AD for the second process $\times$ AI or AD for the cross-extremal dependence). The proposed model can capture five out of these eight cases and has a separate control {on} the joint upper tail and joint lower tail. For example, the process can capture AD in the upper tail and AI in the lower tail, or vice versa. The three cases that the proposed model cannot capture include the cases when {some} of the marginal extremal {dependencies are AI but} the cross-extremal dependence is AD. In other words, when the extremal cross-dependence is AD, both processes must also be AD individually. In practice, this is usually not a major limitation, as the {cross-process} dependence is often found to be weaker in applications than the within-process dependence.
\begin{table}[H]
\centering
\caption{Conditions for asymptotic dependence (AD) in the marginal and cross-extremal dependence structures of our model (\ref{equ:model_i}) within and {across} the different random fields. Conditions for asymptotic independence (AI) are found by symmetry. In the table, the indices are $i, i_1, i_2 = 1,\dots,p$, $i_1 \neq i_2,$ and $j_1, j_2 = 1,\dots,d$.}
\vspace{5pt}
\resizebox{1\textwidth}{!}{
\begin{tabular}{l||l|l}
 & Upper Tail & Lower Tail  \\ \hline \hline
$(X_{i}(\bs s_{j_1}), X_{i}(\bs s_{j_2}))^\top$ 
& $\max(\delta^U, 1-\delta^U)>1/(\alpha_i\gamma_i)$ 
& $\max(\delta^L, 1-\delta^L)>1/\{\alpha_i(1-\gamma_i)\}$  \\
(within-field dependence) &  
& \\ \hline 
$(X_{i_1}(\bs s_{j_1}), X_{i_2}(\bs s_{j_2}))^\top$ 
& $\delta^U>\max\{(\alpha_{i_1}\gamma_{i_1})^{-1},(\alpha_{i_2}\gamma_{i_2})^{-1},1/2\}$ 
& $\delta^L>\max[\{\alpha_{i_1}(1-\gamma_{i_1})\}^{-1},\{\alpha_{i_2}(1-\gamma_{i_2})\}^{-1},1/2]$  \\
(cross-dependence) & & \\ \hline                                                     
\end{tabular}}			

\label{tab:tailproperties}
\end{table}


\subsection{Illustration}\label{subsec:illustration}
We illustrate the flexibility of our proposed model in the bivariate spatial case with $p=2$ by simulation. We focus specifically on the spatial and cross-extremal dependence structures of {$\bs X(\bs s)=(X_1(\bs s), X_2(\bs s))^\top$}, with $X_1(\bs s)$ and $X_2(\bs s)$ defined as in (\ref{equ:model_i}) where we choose an isotropic {exponential correlation} matrix for $W^*_i(\bs s)$, i.e., $\Sigma_i^*:=c_i(H)=\exp(-H/\lambda_i)$, for $i=1, 2$, {$\lambda_i>0$} represent the field-specific spatial range parameters, and $H$ is the distance matrix with $(j_1, j_2)$ element $h_{j_1, j_2} =  ||\bs s_{j_1}-\bs s_{j_2}||$, for $j_1, j_2 = 1,\dots,d$. We specify the elements of the correlation matrix $\Sigma$ by taking $L_{11}=1, L_{21} = \rho_{12},$ and $L_{22} = \sqrt{1-\rho_{12}^2}.$ Recall \S\ref{sec:modelconstruction} for details. Here we collect all parameters in the vector 
$\bs \theta = ( \alpha_1,  \alpha_2, \gamma_1, \gamma_2,  \delta^U, \delta^L, \lambda_1, \lambda_2, \rho_{12})^\top \in (0,\infty)^2\times[0, 1]^4\times(0,\infty)^2\times[-1,1]$. 


\begin{figure}[t!]
\begin{center}
  \begin{minipage}[b]{0.46\linewidth}
    \includegraphics[width=1\linewidth]{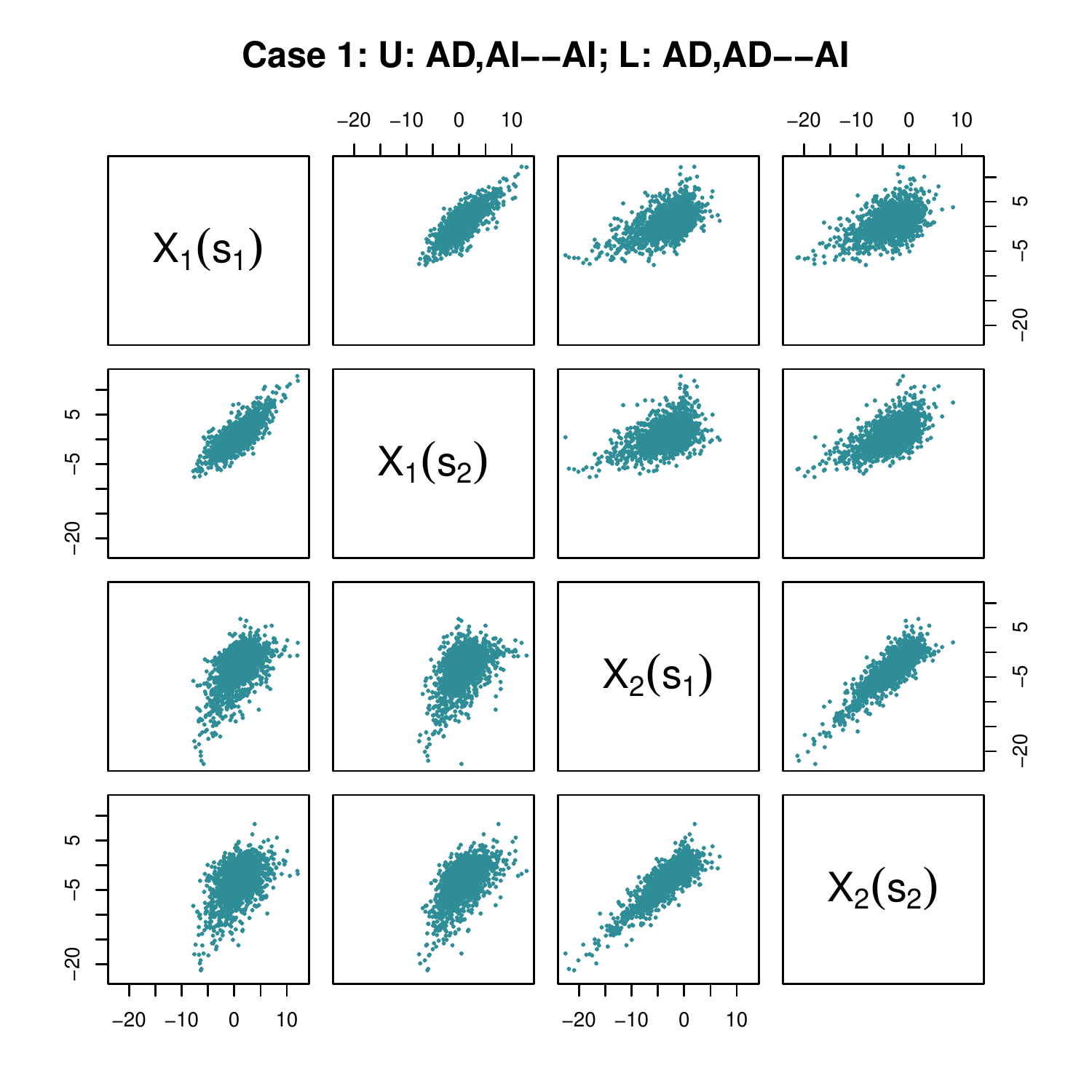} 
  \end{minipage} 
  \begin{minipage}[b]{0.46\linewidth}
    \includegraphics[width=1\linewidth]{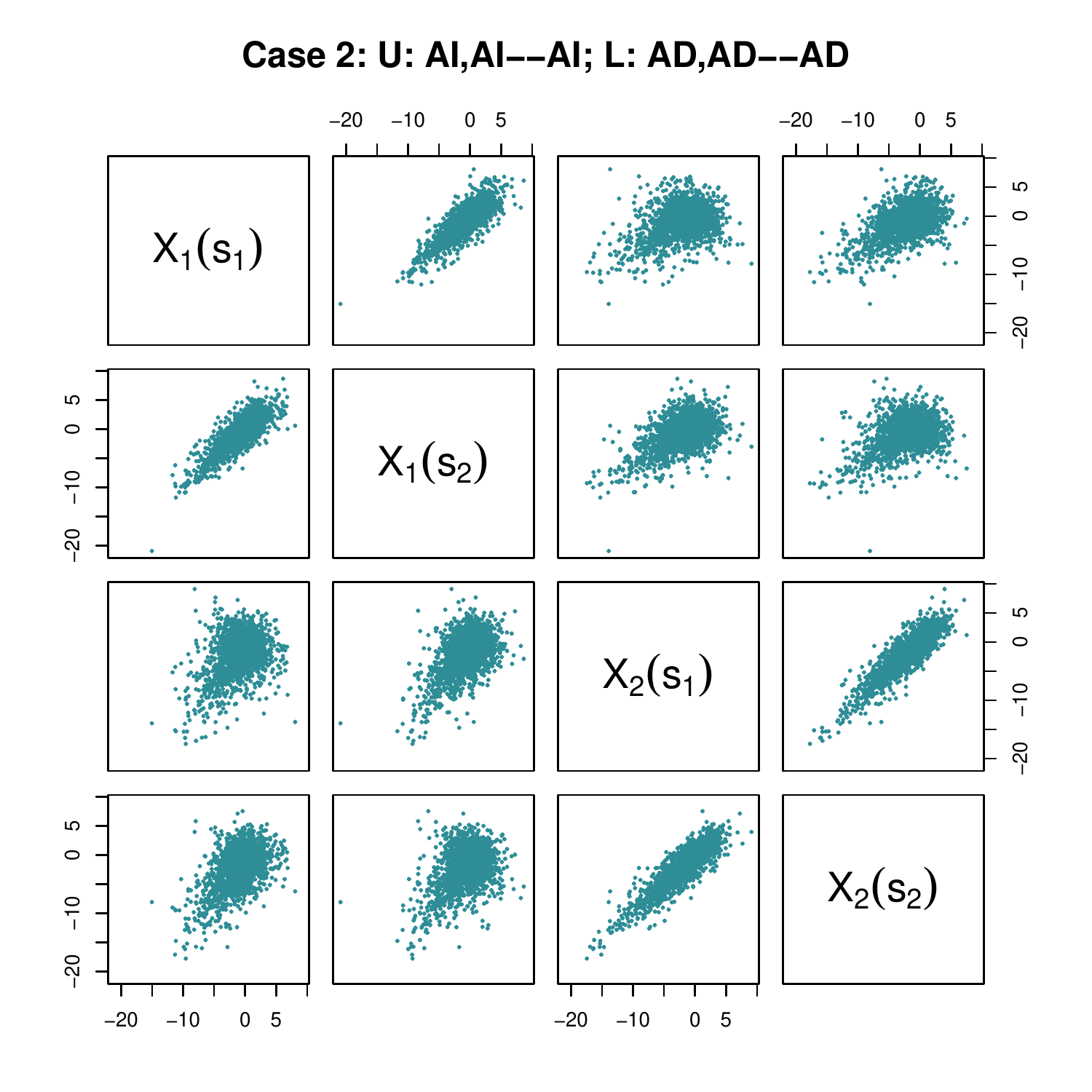} 
  \end{minipage} 
    \hfill
  \begin{minipage}[b]{0.46\linewidth}
    \includegraphics[width=1\linewidth]{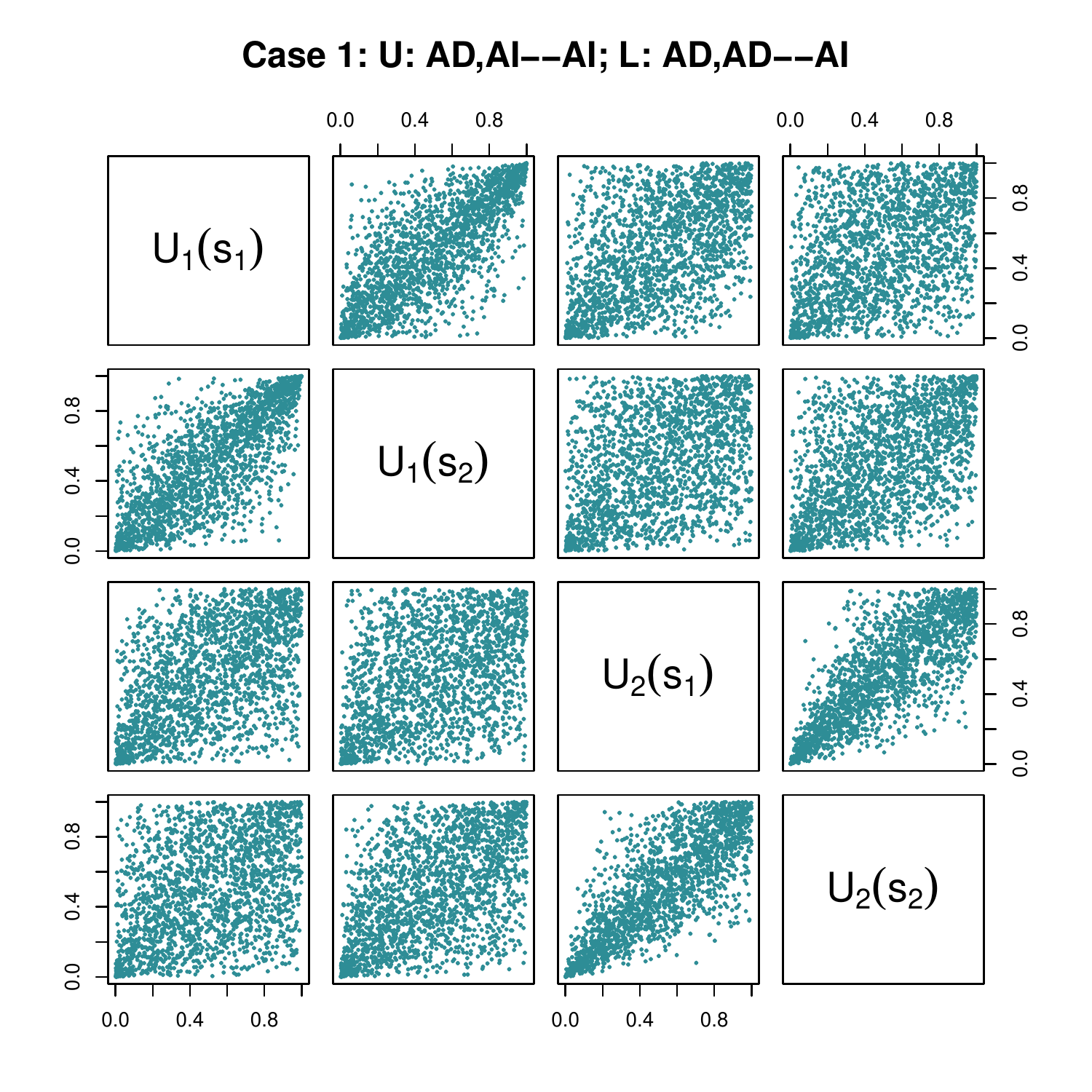} 
  \end{minipage} 
  \begin{minipage}[b]{0.46\linewidth}
    \includegraphics[width=1\linewidth]{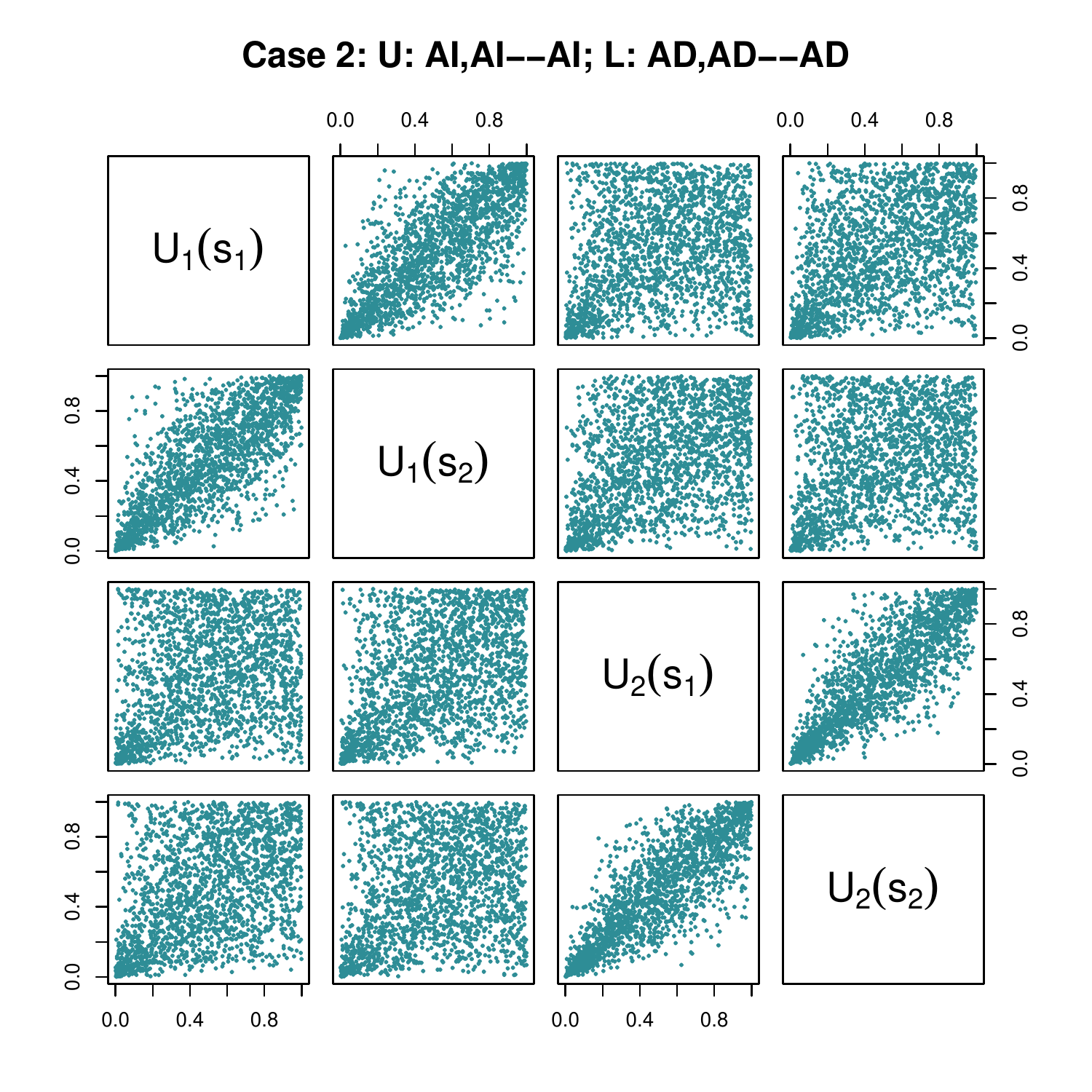} 
  \end{minipage} 
\end{center}
\caption{Simulation results for {Cases 1 and 2} (left and right columns, respectively). For each field, we generated $n = 2000$ replicates {at} two random locations within {the} unit square. The simulated data are plotted both {on} the original scale {of $\bs X$} ({top} row) and {the} uniform scale ({bottom} row). In Case 1,  the upper tail {spatial} extremal {dependence types (i.e., for each of the two individual fields)} are AD and AI, {respectively,} and the upper tail cross-extremal dependence is AI. We write it as U: AD, AI--AI. We adopt a similar notation for the lower tail case. All settings can be read from the titles. Precisely, the parameter settings for both cases are 
$\bs \theta = (\alpha_1 = 4, \alpha_2 = 5, \gamma_1 = 0.6, \gamma_2 = 0.2, \delta^U = 0.4, \delta^L = 0.7, \lambda_1 = 0.7,\lambda_2 = 0.3, \rho_{12} = 0.7)^\top$ (Case 1) and $\bs \theta = (\alpha_1 = 4, \alpha_2 = 5, \gamma_1 = 0.4, \gamma_2 = 0.3, \delta^U = 0.4, \delta^L = 0.8, \lambda_1 = 0.7,\lambda_2 = 0.3, \rho_{12} = -1)^\top$ (Case 2).
}
\label{fig:simu4} 
\end{figure}
Figure \ref{fig:simu4} shows realizations from our model in two scenarios. For each field, we {generate} $n = 2000$ replicates at two random locations within a unit square. The simulated data are plotted both {on} the original scale {of $\bs X$} ({top} row) and {the} uniform scale ({bottom} row). In Case 1 {(left column)},  the upper tail {spatial extremal dependence types (i.e., for each of the two individual fields)} are AD and AI, {respectively,} while the upper tail cross-extremal dependence is AI, which we write concisely in the title of the corresponding panel as U: AD, AI--AI. The lower tail {structure} in Case 1 is chosen to be AD {for} each {individual} marginal field, and AI for the cross-dependence structure, i.e., we write it as L: AD, AD--AI. Similarly, using the same notation, the extremal dependence types of Case 2 {(right column)} are U: AI, AI--AI and L: AD, AD--AD. In both cases, we {fix} the following parameters:  $\alpha_1 = 4, \alpha_2 = 5, \lambda_1 = 0.7$, and $\lambda_2 = 0.3$. However, in Case 1, we {further} specify, $\gamma_1 = 0.6, \gamma_2 = 0.2, \delta^U = 0.4, \delta^L = 0.7$ and $\rho_{12} = 0.7$, while in Case 2, $\gamma_1 = 0.4, \gamma_2 = 0.3, \delta^U = 0.4, \delta^L = 0.8$, and $\rho_{12} = -1$. The simulation shows the flexibility of our model to capture tail asymmetry and to separately control the spatial and cross-extremal dependences in both upper and lower tails.

                    \section{Inference}\label{sec:inference}
\subsection{Marginal modeling}
Let $(\boldsymbol{Y}^\top_{1,1},\dots,\boldsymbol{Y}^\top_{p,1})^\top,\ldots,(\boldsymbol{Y}^\top_{1,n},\dots,\boldsymbol{Y}^\top_{p,n})^\top$ denote $n$ independent copies from a random vector $\boldsymbol{Y}=(\bs Y_1^\top,\dots,\bs Y_p^\top)^\top$ that shares the same copula (\ref{eq:CU4}) as the vector $\boldsymbol{X}$ in (\ref{equ:model_i}) but possesses potentially different marginal distributions $F^Y_{1},\dots, F^Y_{p}$, here assumed to be {continuous and} the same across space, though it can be easily generalized to the case where each spatial process has non-stationary margins. {We write} $\bs Y_{i,k}=(Y_{i1,k},\dots,Y_{id,k})^\top$ for $i=1,\dots,p$ and $k=1,\dots,n.$ 
{The} joint distribution of $\boldsymbol{Y}$ can be expressed as 
\begin{align}
\notag F^{\bs Y}(\bs y_1,\dots, \bs y_p)&=\pr(\bs Y_1\leq \bs y_1,\dots, \bs Y_p\leq \bs y_p)\\
\notag &=F^{\boldsymbol{X}}[(F^X_{1})^{-1}\{F^Y_{1}(\bs y_{1})\},\dots, (F^X_{p})^{-1}\{F^Y_{p}(\bs y_{p})\}]=C^{\bs X}\{F^Y_{1}(\bs y_{1}),\dots, F^Y_{p}(\bs y_{p})\},
\end{align}
where the functions $(F^X_{i})^{-1}$ and $F^Y_{i}, i = 1,\ldots,p,$ are applied componentwise. To estimate the copula $C^{\bs X}$ from an observed random sample $(\bs y_{1,1}^\top, \dots, \bs y_{p,1}^\top)^\top,\ldots,(\bs y_{1,n}^\top, \dots, \bs y_{p,n}^\top)^\top$, where $(\bs y_{1,k}^\top, \dots, \bs y_{p,k}^\top)^\top=(y_{11,k},\dots,y_{1d,k},\dots,y_{p1,k},\dots,y_{pd,k})^\top, k = 1, \ldots, n$, we first need to transform the data to the standard uniform scale. To achieve this goal, we may either estimate $F^{Y}_i$ using a parametric model, or more simply {using} the empirical distribution function given as
$$\hat F^Y_{i}(y) = {1\over nd+1}\sum_{j=1}^d\sum_{k=1}^n I(y_{ij,k}\leq y), \quad {i=1,\dots,p}.$$
Note that this formula needs to be adjusted when the data have different marginal distributions at different sites. Then we define pseudo-uniform scores using their corresponding ranks {(computed among all $d$ spatial observations and $n$ independent replicates, but separately for each field $i=1,\ldots,p$), i.e., } $$u_{ij,k}=\hat F^Y_{i}(y_{ij,k})={{\rm rank}(y_{ij,k})\over nd+1}, \quad i=1,\dots,p,\quad j = 1,\ldots,d{,\quad k=1,\ldots,n}.$$
The joint distribution of the proposed model cannot be obtained in the closed form. However, by conditioning on the latent factors and model parameters, the copula can be expressed analytically in terms of the Gaussian copula{, which is computationally convenient}. This suggests adopting {a Bayesian inference approach based on Markov chain Monte Carlo (MCMC) methods}, by treating the unobserved common factors as latent variables to be updated at each iteration of the MCMC algorithm.

\subsection{Bayesian inference}\label{sec:bayesian inference}
Conditioning on the model parameter {vector} $\bs \theta$ and latent variables $\bs R = (R^U_{0}, R^L_{0}, R^U_{1}, R^L_{1}, \dots, R^U_{p}, R^L_{p})^\top$, we have that the copula $C^{\bs X}$ of uniformly transformed observations from model (\ref{equ:model_i}) is
{\begin{align}
\notag C^{\bs X}(\bs u_1,\dots, \bs u_p\mid \bs \theta, \bs R) =& \pr (\bs U_1\leq\bs u_1, \dots,\bs U_p\leq\bs u_p\mid \bs \theta, \bs R)\\
	\notag=&\Pr(\bs X_1\leq (F^X_1)^{-1}(\bs u_1),\dots, \bs X_p\leq (F^X_p)^{-1}(\bs u_p)\mid \bs \theta, \bs R)\\
	=& \bs \Phi_{pd}\Big(
	\Phi^{-1}[F_W(\bs w_1)],\dots,\Phi^{-1}[F_W(\bs w_p)]; \bs 0_{pd}, \Sigma_{pd\times pd} \Big),\label{eq:condC}
\end{align}}
where 
{$\bs w_{i} =(F^X_i)^{-1}(\bs u_i)- \alpha_i\big\{ \gamma_i[\delta^UR_0^U+(1-\delta^U)R_i^U]-(1-\gamma_i)[\delta^LR_0^L+(1-\delta^L)R_i^L]\big\}$}, for $i=1,\dots,p$, $\Phi^{-1}$ is the univariate Gaussian quantile function and $\Phi_{pd}(\cdot; \bs 0_{pd}, \Sigma_{pd\times pd})$ denotes the multivariate Gaussian distribution in dimension $pd$ with mean zero and correlation matrix $\Sigma$.
Therefore, with $n$ independent realizations of $\bs u = (\bs u_1, \dots,\bs u_p)^\top$, we hereby perform Bayesian inference of our model by jointly updating $\bs \theta$ and the $n$ independent copies of $\bs R$. 
To simplify the sampling scheme, we first transform {$\bs \theta$ and $\bs R$ into new variables that are supported on the whole real line.} In particular, we reparametrize $\bs \theta$ as $\bs \theta^* = g(\bs \theta)$, such that ${\rm supp}(\bs \theta^*)=\Real^M$, where {$M=9$} (when no parameter is fixed) is the dimension of the parameter space. Similarly, we reparametrize $\bs R$ as $\bs R^* = h(\bs R)$, such that ${\rm supp}(\bs R^*)=\Real^Q$, where $Q = 2(p+1)n$ is the dimension of the latent variable space. We use here log-transforms for positive variables or parameters, and (possibly rescaled) logit transforms for bounded parameters. 
The joint posterior density of $\bs \theta^*$ and {the} transformed latent variables $\bs R^*$, for a single replicate, {may} then be expressed as
{\begin{align}\notag
\pi(\bs \theta^*, \bs R^* \mid \bs u)
&\propto \pi(\bs u\mid \bs \theta^*, \bs R^*)\times\pi(\bs \theta^*, \bs R^*)\\
\notag&\propto \pi(\bs u\mid \bs \theta, \bs R)\times\pi(\bs R^*)\times\pi(\bs \theta^*),
\end{align}}
where $\bs u = (\bs u_1^\top, \ldots, \bs u_p^\top)^\top$ {are the observed data,} and the {density $\pi(\bs u\mid \bs \theta^*, \bs R^*)$} can be derived {by differentiating (\ref{eq:condC}) with respect to $\bs u$}. The density {$\pi(\bs{R^*})=\pi(\bs{R})/{\partial h(\bs R)\over\partial\bs{R}}$}, derived from the change-of-variable formula, {can be computed from $\pi(\bs{R})$}, which is a product of standard exponential densities (as defined in the proposed model). The term $\pi(\bs \theta^*)$ is the prior density of $\bs \theta^*$ and can be specified similarly as $\pi(\bs{\theta^*})=\pi(\bs{\theta})/{\partial g(\bs \theta)\over\partial\bs{\theta}}$, where $\pi(\bs{\theta})$ is the prior density of {$\bs\theta$} on its original scale. {With $n$ independent replicates, the densities $\pi(\bs u\mid \bs \theta^*, \bs R^*)$ (i.e., the likelihood function) and $\pi(\bs{R^*})$ are replaced by products of $n$ similarly defined terms, and the joint posterior density has to be modified accordingly. Precisely, the likelihood function with $n$ replicates is}
{\begin{align}
	\notag \pi( \{\bs u_k\}_{k=1}^n \mid \bs \theta,  \{\bs R_k\}_{k=1}^n)=&
	\prod_{k=1}^n
	\pi(\bs u_{1,k}, \dots,\bs u_{p,k}\mid  \bs\theta, \bs R_{k}),
\end{align}}
where $\bs u_k = (\bs  u_{1,k}^\top, \ldots, \bs  u_{p,k}^\top)^\top$, $\bs u_{i,k} = (u_{i1,k}^\top, \ldots, u_{id,k}^\top)^\top$, {denotes} the {$k$-th uniform data} replicate associated with the {$k$-th independent} latent random factor {$\bs R_k$}, and the log-likelihood {thus becomes}
{\begin{align}
\notag\ell( \{\bs u_k\}_{k=1}^n\mid \bs \theta,  \{\bs R_k\}_{k=1}^n)
 = & \log\pi( \{\bs u_k\}_{k=1}^n\mid \bs \theta,  \{\bs R_k\}_{k=1}^n)\\
          =&\sum_{k=1}^n\log\pi(\bs u_{1,k}, \dots,\bs u_{p,k}\mid \bs\theta, \bs R_{k}) \\
\notag=& 
	\sum_{k=1}^n\Big[
	\log \phi_{pd}\big(\Phi^{-1}[F_W(\bs w_{1,k})],\dots,\Phi^{-1}[F_W(\bs w_{p,k})]\mid \bs\theta, \bs R_{k}\big)\\
\notag+&
         \sum_{i=1}^p\sum_{j=1}^d
         \log f_W(w_{ij,k})-\log f_i^X[(F^X_i)^{-1}(u_{ij,k})]-\log\phi(\Phi^{-1}[F_W(w_{ij,k})])
	\Big],
\end{align}}
{where} $\bs w_{i,k} = (w_{i1,k}, \ldots, w_{id,k})^\top$, $i=1,\ldots p,$ {denotes} the $k$-th replicate of $\bs w_i$. {Using the expressions above, the overall joint log posterior density can thus easily be computed, up to an additive constant, as
$$\log\pi(\bs \theta^*, \{\bs R^*_k\}_{k=1}^n \mid \{\bs u_k\}_{k=1}^n)\equiv \ell( \{\bs u_k\}_{k=1}^n\mid \bs \theta,  \{\bs R_k\}_{k=1}^n)+\sum_{k=1}^n\log\pi(\bs R^*_k)+\log\pi(\bs\theta^*).$$}


\subsection{{MCMC algorithm}}\label{sec:mcmc algorithm}
We hereby introduce our {proposed MCMC algorithm customized to fit model \eqref{equ:model_i}}. 
First, we {present} the general Metropolis--Hastings algorithm {procedure and then describe proposal distributions that are designed to sample efficiently from the joint posterior distribution of hyperparameters and latent variables}. 
Let $\pi(\bs z)$ be the target density {(e.g., the posterior density)} of an MCMC sampler where $\bs z$ is an $m$-dimensional vector {of variables we need to infer}. 
The basic Metropolis--Hastings algorithm is constructed by {sampling candidate values $\bs z'$ according to a proposal distribution that has density} $\bs z'\sim q(\cdot\mid \bs z)$ (also known as the Markov kernel), and to accept {the} proposed $\bs z'$ with probability $\alpha = \min\Big\{1, {\pi(\bs z') q(\bs z\mid \bs z') \over \pi(\bs z) q(\bs z'\mid \bs z) }\Big\}$.
Based on the detailed balance property, if the resulting Markov chain $\{\bs z^{(t)}\}_{t=1,\dots,N}$ ({where} $N$ is the total number of iterations) is irreducible \citep{robert2013monte}, convergence to the target distribution $\pi(\bs z)$ is guaranteed from this acceptance-rejection procedure. The choice of suitable proposal distributions is crucial for obtaining good performances (i.e., good mixing of Markov chains and fast convergence). We now describe {our approach} to {efficiently sample candidate values of hyperparameters and high-dimensional latent variables from their posterior density} in our case. Our proposed methodology relies on three features:
\begin{enumerate}[(i)]
\item {Random walk proposal for hyperparameters}\label{sec: rw}. {To update the vector $\bs \theta^*\in\mathbb R^M$, we} choose a random walk proposal {$\bs {\theta^*}'\sim q_{\rm RW}(\cdot\mid\bs \theta^*)$, where $q_{\rm RW}$ is} the density of the Gaussian distribution {$\mathcal{N}_m(\bs \theta^*, \sigma^2_{\rm RW}I_M)$, and $I_M$ is the $M\times M$} identity matrix and $\sigma^2_{\rm RW}>0$ is a tuning parameter that controls the step size.
\item {Metropolis-Adjusted Langevin Algorithm proposal (MALA) for latent variables}\label{sec:Mala}. {To update} the high-dimensional vector of latent variables {$\bs R^*=\{\bs R^*_k\}_{k=1}^n\in\Real^Q$}, we choose a MALA {block} proposal {${\bs R^*}'\sim q_{\rm MALA}(\cdot\mid \bs R^*,\bs\theta^*)$}, which has the density of {$\mathcal{N}_m(\bs R^*+{\sigma^2_{\rm MALA}\over 2}\nabla_{\bs R^*}\log\pi(\bs\theta^*,\bs R^*\mid \bs{u}), \sigma_{\rm MALA}^2 I_{Q})$}, where $\sigma_{\rm MALA}^2>0$ is a tuning parameter {and $I_Q$ is the $Q\times Q$ identity matrix (recall that $Q=2(p+1)n$)}. The MALA proposal was introduced by \cite{roberts1996exponential} and further studied by \cite{roberts1998optimal}. One {key} advantage of MALA {proposals} is the additional information on the gradient {$\nabla_{\bs R^*}\log\pi(\bs\theta^*\bs R^*\mid \bs{u})$, which allows} exploring {high-dimensional parameter spaces} more efficiently.
\item {Adaptive strategy}\label{sec:adap}. As suggested in the above proposals, the tuning {parameters $\sigma_{\rm RW}^2$ and $\sigma_{\rm MALA}^2$ (denoted generically below by $\sigma^2$) directly affect} the step size and {Metropolis--Hastings} acceptance rate, $a$. {Let $N$ be the total length of the chain and $N_b$ be the number of iterations during the initial burn-in period. For $t\leq N_b$, we tune $\sigma^2$ adaptively by updating its value every $N_0<N_b$ iterations as} 
\begin{equation}\label{equ:update}
\sigma \mapsto \sigma' : = \sigma \exp\Big({a_0-a^* \over b}\Big),
\end{equation}
where $b>0$ is a scaling parameter, $a^*$ is the target acceptance rate, and $a_0$ is the current acceptance rate for every $N_0$ iterations.
\cite{roberts1998optimal} suggested that the optimal acceptance rate for the MALA is 0.574, {while \cite{roberts1997weak} concluded that it is around 0.234 for the random walk Metropolis--Hastings} sampler. Even though the settings in the above references are different from the present {case}, these rates can still be considered {as helpful guidelines}. Thus, similar to \cite{yadav2021spatial,yadav2021flexible}, we set $a^*=0.574$ for MALA proposals and $a^* = 0.234$ for random walk proposals.
\end{enumerate}
The developed algorithm is summarized in Algorithm \ref{algo1}.
\begin{algorithm}[h!]
	\SetKwInOut{Input}{Input}
	\SetKwInOut{Output}{Output}
	\SetKwInOut{Parameter}{Parameter}
	\SetKw{Return}{Return}
	\Input{Data $\bs u =  \{\bs u_k\}_{k=1}^n$; Initial parameters $\{\bs \theta^{*(0)}, \bs R^{*(0)}\}$, where $\bs R^{*(0)}= \{\bs R^{*(0)}_k\}_{k=1}^n$.}
	\Output{MCMC chain $\{\bs \theta^{*(t)}, \bs R^{*(t)}\}_{t=1,\dots,N}$, where $\bs R^{*(t)}= \{\bs R^{*(t)}_k\}_{k=1}^n$.}
	\For{$t\leftarrow 1$ \KwTo $N$}{
\emph{\tcp{Block 1: updating $\bs \theta^*$ with RW proposal}}
	$\bs \theta^{*(p)}\sim q_{\rm RW}(\cdot \mid \bs \theta^{*(t-1)})$\;
		$\alpha_{\rm RW}(\bs \theta^{*(t-1)}, \bs \theta^{*(p)}) \leftarrow \min
		\Big\{1,{ \pi(\bs \theta^{*(p)}, \bs R^{*(t-1)}\mid \bs{u})\over \pi(\bs \theta^{*(t-1)}, \bs R^{*(t-1)}\mid \bs{u})} \Big\}$; Draw $U\sim U(0,1)$\;
	\eIf{$\alpha_{\rm RW}(\bs \theta^{*(t-1)}, \bs \theta^{*(p)})\leq  U$}{$\bs\theta^{*(t)} \leftarrow \bs \theta^{*(p)}$}
	{$\bs \theta^{*(t)} \leftarrow \bs \theta^{*(t-1)}$ }
 \emph{\tcp{Block 2: updating latent variables $\bs R^*$ with MALA proposal}}
$\bs R^{*(p)}\sim q_{\rm MALA}(\cdot \mid \bs \theta^{*(t)}, \bs R^{*(t-1)})$; 

		$\alpha_{\rm MALA}(\bs R^{*(t-1)}, \bs R^{*(p)}) \leftarrow \min\Big\{1,{ 
		\pi(\bs \theta^{*(t)}, \bs R^{*(p)}\mid \bs{u})q_{\rm MALA}(\bs R^{*(t-1)}\mid \bs R^{*(p)},\bs \theta^{*(t)})
		\over 
		\pi(\bs \theta^{*(t)}, \bs R^{*(t-1)}\mid \bs{u})q_{\rm MALA}(\bs R^{*(p)}\mid \bs R^{*(t-1)}, \bs \theta^{*(t)})
		} \Big\}$; Draw $V\sim U(0,1)$\;
		\eIf{$\alpha_{\rm MALA}(\bs R^{*(t-1)}, \bs R^{*(p)})\leq V$}{$\bs R^{*(t)} \leftarrow \bs R^{*(p)}$}
		{ $\bs R^{*(t)} \leftarrow \bs R^{*(t-1)}$}
		 \emph{\tcp{Adaptive strategy}}
\If{$t<N_b$ and $t\;({\rm mod} \;N_0) = 0$}{update both $\sigma_{RM}$ and $\sigma_{\rm MALA}$ according to (\ref{equ:update});}
	}
	\Return{$\{\bs \theta^{*(t)}, \bs R^{*(t)}\}_{t=1,\dots,N}$}\;
\caption{MCMC Algorithm.}\label{algo1}
\end{algorithm}

\section{Simulation study}\label{sec:simulation}
\subsection{{Simulation setting and two-step estimation approach}}
In the simulation study, we {mimic our real data application and} generate $n = 300$ {independent replicates at $d=25$ randomly sampled locations on the unit square $[0,1]^2$} using the {bivariate ($p=2$) spatial} model specified in \S\ref{subsec:illustration}. Here the true parameters are set to $\alpha_1 = 4$, $\alpha_2 = 4$, $\gamma_1 = 0.4$, $\gamma_2 = 0.6$, $\delta^U = 0.8$, $\delta^L = 0.6$, $\lambda_1 = 0.6$, $\lambda_2 = 0.3$, and $\rho_{12} = -0.7${, such that the} resulting marginal and cross-extremal dependences of the upper and lower tails are as follows: U: AD, AD--AD and L: AD, AI--AI {using the notation of \S\ref{subsec:illustration}}. 

Although all parameters of model \eqref{equ:model_i} are theoretically identifiable, some may be weakly identifiable in practice as they all control {various bulk and tail} dependence characteristics. This means that all parameters may be difficult to estimate simultaneously. {This is particularly challenging in relatively low sample sizes, and especially for estimating the parameters $\delta^U$ and $\delta^L$, which are mostly related to the cross-process tail structure.} Therefore, to simplify inference, we suggest {adopting a two-step inference approach, whereby} $\alpha_1, \alpha_2, \delta^U$, and $\delta^L$ {are initially fixed} to some reasonable candidate values{, and all other five parameters are then estimated using the Bayesian algorithm outlined in \S\ref{sec:inference}. The different candidate models can finally be compared using various visual or more objective model diagnostics.} Here, in this simulation, we only illustrate the second step for simplicity, i.e., we show that our Bayesian approach works well {when} $\alpha_1, \alpha_2, \delta^U, \delta^L$ {are} fixed to their true values. Concerning the priors for {$\bs\theta=(\gamma_1,\gamma_2,\lambda_1,\lambda_2,\rho_{12})^\top$}, we assign Exp(1) priors for all hyperparameters that have a domain at the positive real line, while we use uniform priors for parameters with bounded domains. {Furthermore,  we set the number of burn-in iterations to $N_b = 10^5$, the total number of iterations to $N =5\times10^5$, and we monitor convergence of Markov chains through traceplots.}

\subsection{Extremal dependence diagnostics}
To illustrate the performance of our {Bayesian} inference approach, we compare the {bivariate} extremal dependence {measure}, $\chi$, estimated either empirically {from the data} or based on the fitted model, to the true $\chi$ value {obtained by simulating from \eqref{equ:model_i} with a very large number of replicates}. Let ${\chi}_{11}^U(h, u)$, ${\chi}_{22}^U(h, u)$, and ${\chi}_{12}^U(h, u)$ denote the true sub-asymptotic {upper tail dependence measure $\chi$} between two variables from the first field, second field, and both fields, respectively, at distance $h$ and threshold level $u\in (0,1)$, i.e., {$\chi_{i_1, i_2}^U(h,u) = \Pr(U_{i_1}(\bs s_{j_1})>u\mid U_{i_2}(\bs s_{j_2})>u)$} for $i_1, i_2=1,2$, and {$h=\|\bs s_{j_1}-\bs s_{j_2}\|$, $\bs s_{j_1},\bs s_{j_2}\in \mathcal{S}$, where $U_i(\bs s)=F_i^X\{X_i(\bs s)\}$ denotes the $i$-th random field transformed to the uniform scale through its marginal distribution $F_i^X$. By symmetry, a similar quantity, denoted $\chi_{i_1, i_2}^L(h,u)$, can be defined for the lower tail.} We display the results by plotting them as a function of distance $h$ and threshold $u$. 

\subsubsection{Performance for fixed threshold}

\begin{figure}[t!]
	\centering
	\includegraphics[width=0.94\linewidth]{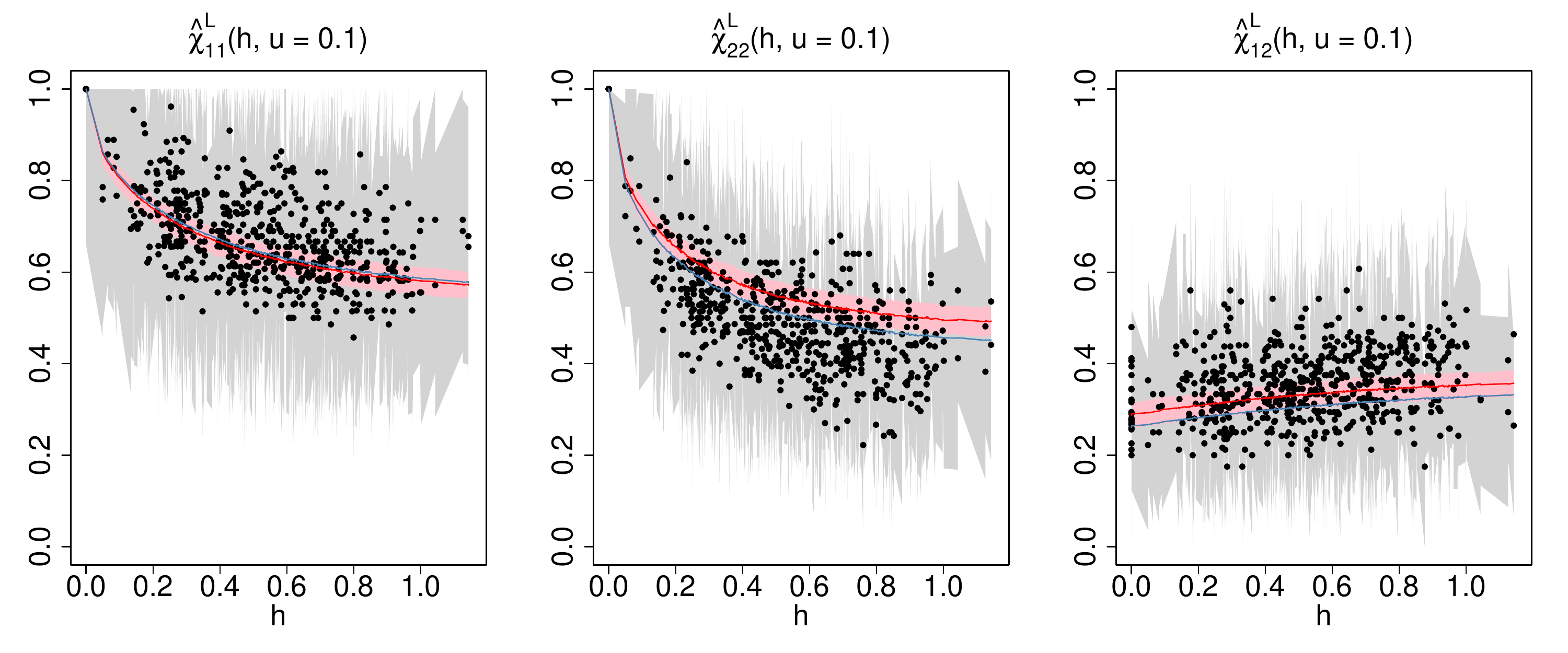}
	\includegraphics[width=0.94\linewidth]{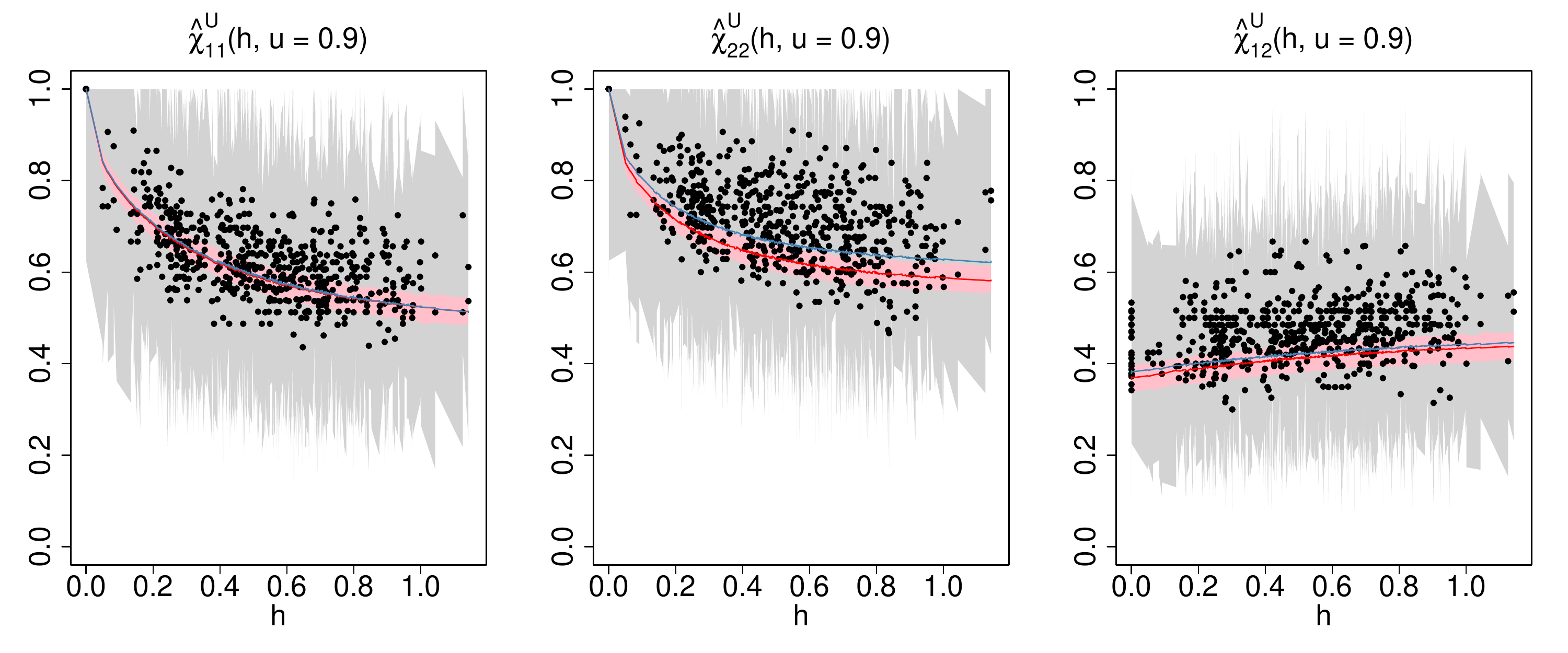}
	\caption{{Empirical estimates (black dots), model estimates (red lines) and true values (blue lines) of ${\chi}_{11}^L(h, u)$, ${\chi}_{22}^L(h, u)$, ${\chi}_{12}^L(h, u)$ (top row, left to right) for fixed threshold $u=0.1$, and ${\chi}_{11}^U(h, u)$, ${\chi}_{22}^U(h, u)$, ${\chi}_{12}^U(h, u)$ (bottom row, left to right) for fixed threshold $u=0.9$, as a function of distance $h$. The gray shadow envelopes are pointwise 95\% confidence intervals for the empirical estimates. The pink shadow envelopes are 95\% credible intervals sampled from the post-burn-in MCMC chains.}}
	\label{fig:simu_01}
\end{figure}

{We first display the spatial extremal dependence estimates with respect to distance for the fixed thresholds $u=0.1$ and $u=0.9$ for the lower and upper tails, respectively. Figure \ref{fig:simu_01} reports the results. Specifically, we plot the empirical estimates (black dots), model-based estimates (red lines) and true values (blue lines) of ${\chi}_{11}^L(h, u)$, ${\chi}_{22}^L(h, u)$, ${\chi}_{12}^L(h, u)$ for fixed threshold $u=0.1$, and ${\chi}_{11}^U(h, u)$, ${\chi}_{22}^U(h, u)$, ${\chi}_{12}^U(h, u)$ for fixed threshold $u=0.9$, as a function of distance $h$ between sites.} The envelopes displayed as gray shadows are {pointwise} 95\% confidence intervals for the empirical estimates, and the envelopes displayed as pink shadows are 95\% credible intervals for model-based estimates, obtained from the post-burn-in MCMC samples. We can see that while empirical estimates are very variable, our inference procedure performs very well at recovering the true {tail} dependence characteristics, with low uncertainty.

\subsubsection{Performance for fixed distance}
\begin{figure}[t!]
	\centering
	\includegraphics[width=0.94\linewidth]{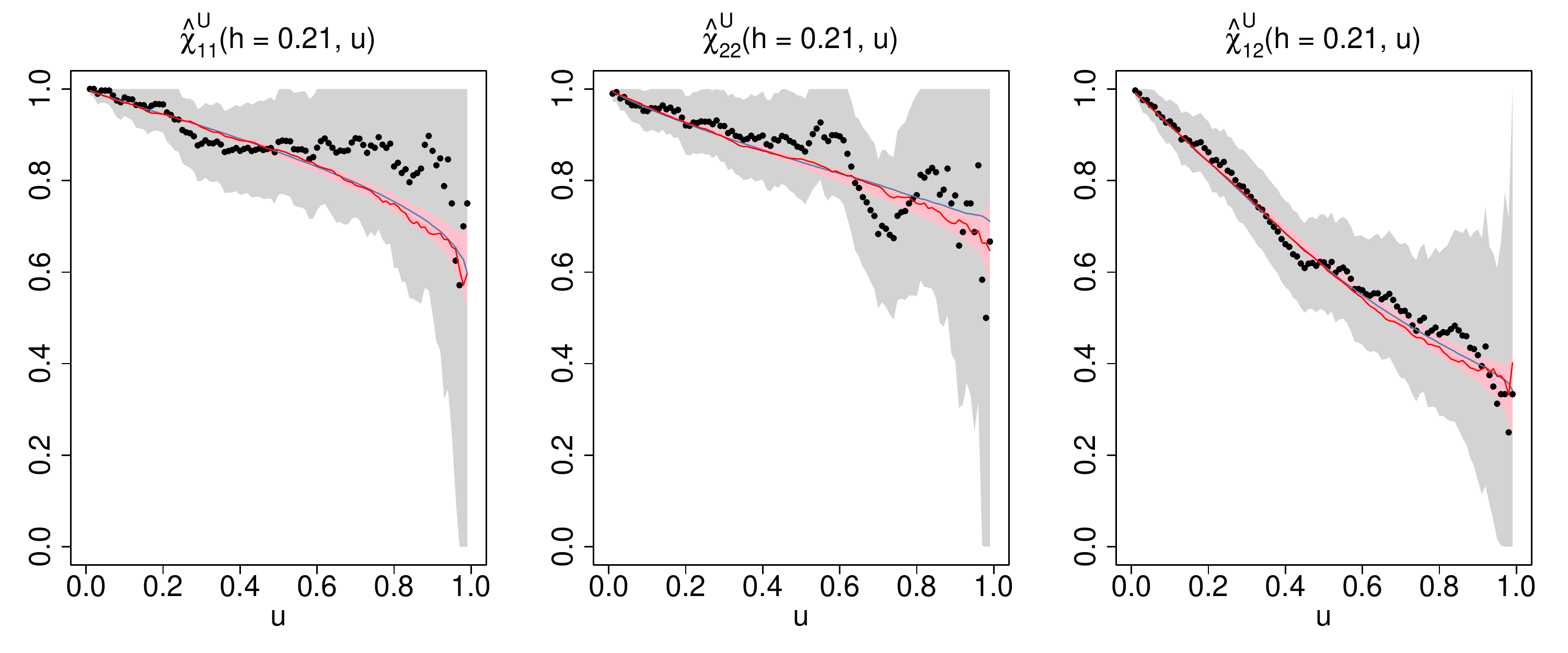}
	\includegraphics[width=0.94\linewidth]{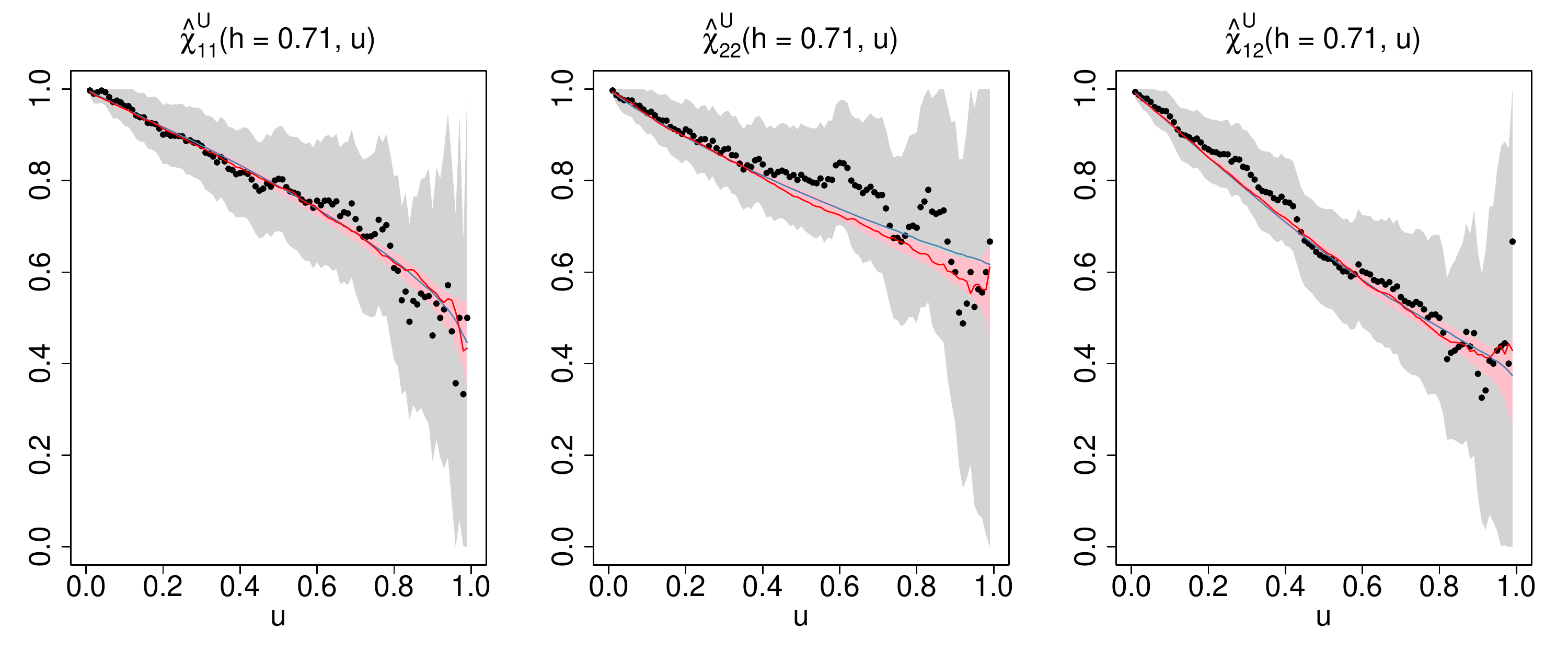}
	\includegraphics[width=0.94\linewidth]{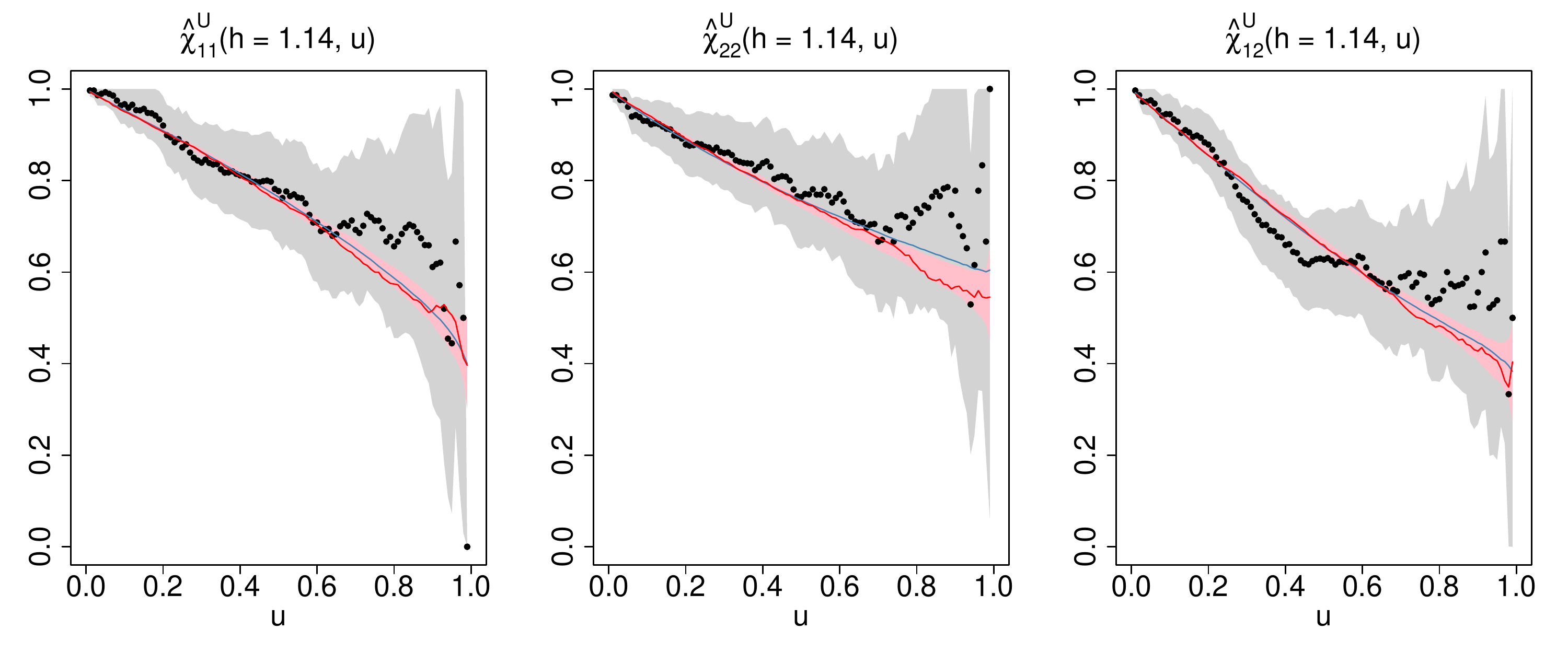}
	\caption{{Empirical estimates (black dots), model estimates (red lines) and true values (blue lines) of ${\chi}_{11}^U(h, u)$, ${\chi}_{22}^U(h, u)$, ${\chi}_{12}^U(h, u)$ (left to right) as a function of the threshold $u$, for fixed distance $h=0.21,0.71,1.14$ (top to bottom), i.e., short, medium and long distance, respectively. The gray shadow envelopes are pointwise 95\% confidence intervals for the empirical estimates. The pink shadow envelopes are 95\% credible intervals sampled from the post-burn-in MCMC chains.}}
	\label{fig:simu_up_short}
\end{figure}

{We then display estimates of the upper tail extremal dependence measures ${\chi}_{11}^U(h, u)$, ${\chi}_{22}^U(h, u)$, ${\chi}_{12}^U(h, u)$ estimates with respect to the threshold level $u$ for the upper tail and fixed distance $h$. Figure \ref{fig:simu_up_short} reports the results for short, median, and long distances (recall that the possible distances in the unit square ranges approximately from zero to $1.41$). Results for the lower tail at different distances are given in the Supplementary Material.} 
The conclusion {is the same as} for the fixed threshold case. From our simulation experiments, we can see that the proposed model can describe and capture the {within-process} and cross-extremal spatial dependences flexibly among different {distances considered} for both the upper and lower tails, and that our inference procedure works well.

                                   \section{Real data application to bivariate temperature extremes}\label{sec:application}
\subsection{Context, data description and pre-processing}
The state of Alabama, in the southeastern US, has humid subtropical climate and with its proximity to the Gulf of Mexico, {it is} also subject to extreme weather events like tropical storms{, which include the infamous Hurricanes Camille (1969) and Katrina (2005). Moreover, because} of its warm and humid climate, heat-related health issues in Alabama have been {an important concern for decades} \citep{taylor2000temperature, wang2019estimating, carter2020methods,wang2021heat}. In our {real data application}, we apply the proposed model to study the lower and upper extremal dependence structures of the daily {maximum (TMAX)} and {daily minimum (TMIN)} air temperature data in Alabama, which are closely related to heat waves and tropical nights, which are extreme events {severely affecting} human health. We are interested in modeling jointly the spatial extremal {dependence} within and across the TMAX and TMIN processes, using the proposed model which can {capture and help identify both tail dependence types for the upper and lower tails}. The complete daily dataset is available from {January 1st, 2010, to May 29th, 2021}, giving a total of 3554 observations at $d=18$ monitoring sites in Alabama.

\begin{figure}[t!]
\begin{center}
  \begin{minipage}[b]{0.3\linewidth}
    \includegraphics[width=1\linewidth]{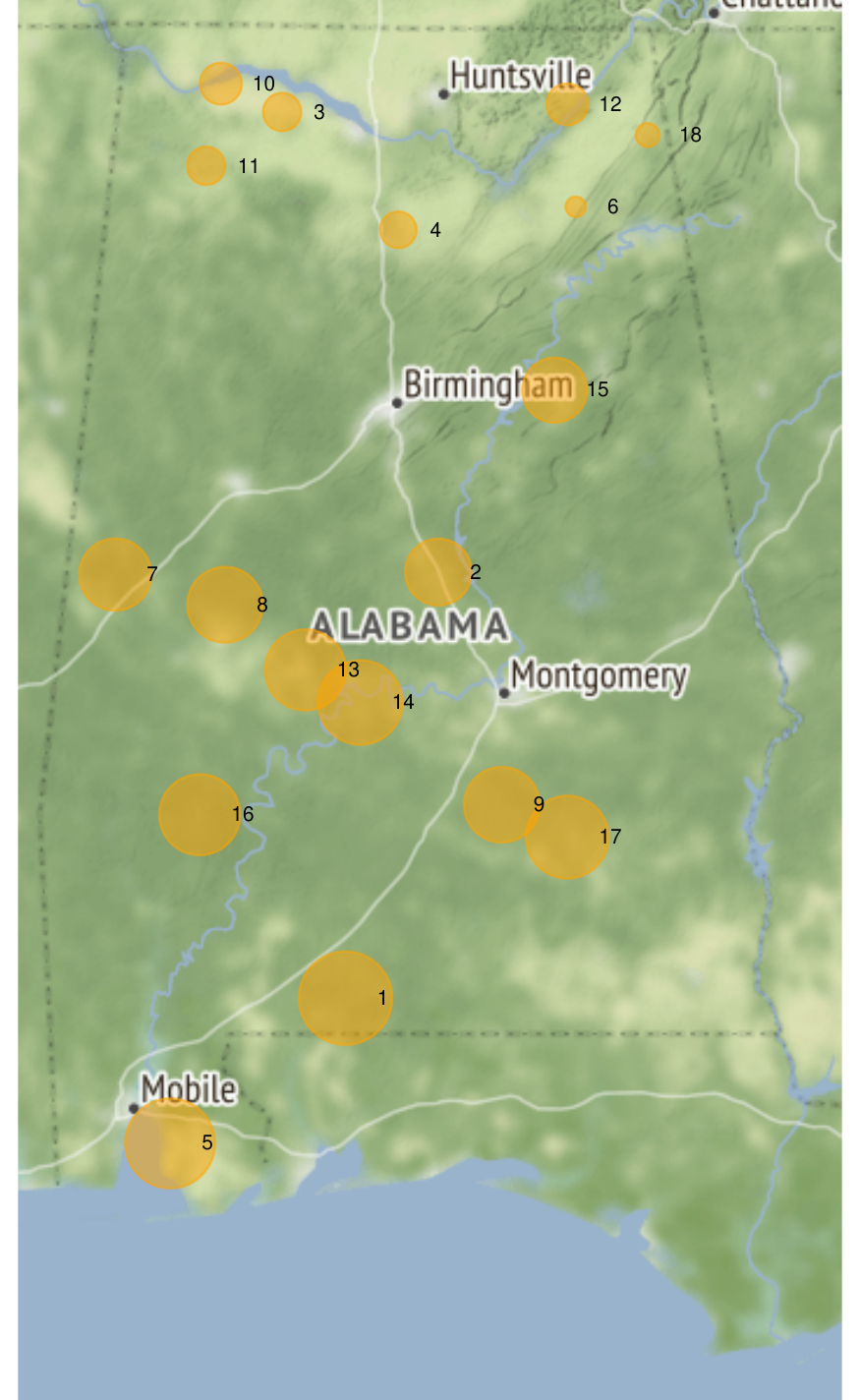} 
  \end{minipage} 
    \begin{minipage}[b]{0.3\linewidth}
    \includegraphics[width=1\linewidth]{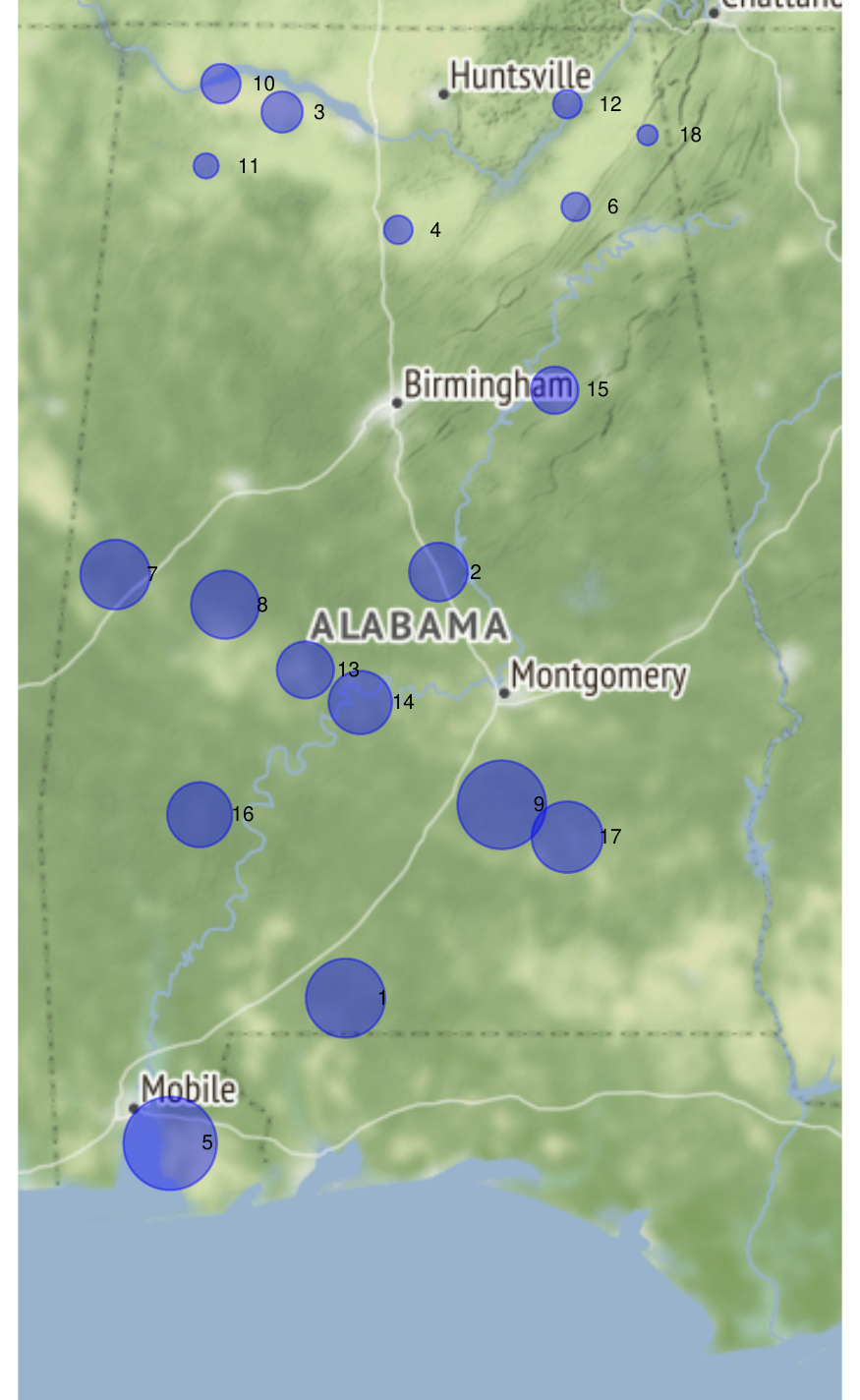} 
  \end{minipage} 
\end{center}
\caption{{Map of Alabama in the Southeastern United States, with the monitoring stations. The size of each circle is proportional to the mean value of TMAX (left, yellow circles) and TMIN (right, blue circles).}}
\label{fig:map} 
\end{figure}

{To focus on a rather extreme year, we here consider data from 2020, which contains the most active Atlantic hurricane season on record with 31 tropical or subtropical cyclones including 14 hurricanes, partly caused by the ``La Ni\~{n}a'' phenomenon that developed in the summer months of 2020. The year 2020 contains $n=271$ daily observations with complete data at the 18 monitoring stations. The maximum distance between two stations is 493.71 km. Figure~\ref{fig:map} displays the map of the study region and monitoring stations, as well as the annual mean value of TMAX and TMIN for 2020. This figure shows that there is a clear correspondence between TMAX and TMIN values, and that high values tend to be spatially clustered and co-located (with larger values close to the South, near the coast), but it is unclear whether, and to what extent, this is due to similar marginal distributions or to the dependence structure.}


%

\begin{figure}[t!]
		\centering
		\includegraphics[width=1\linewidth]{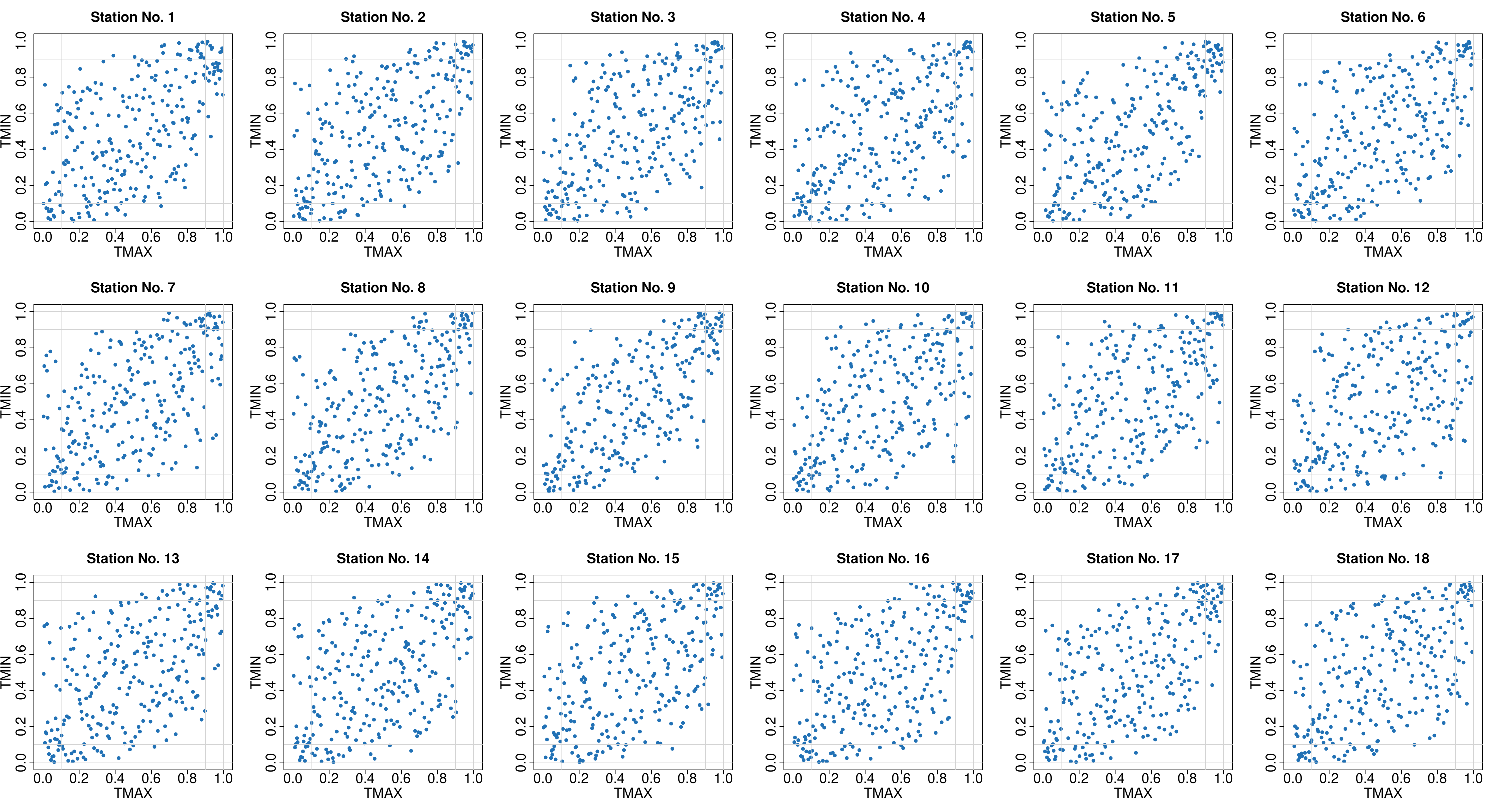}
		\caption{Residuals for each station obtained after removing the seasonalities of the standardized TMAX and TMIN (in uniform scale) for the year 2020.}
		\label{fig:u_resi2020}
\end{figure}

{To disentangle marginal and dependence characteristics of the data, and to fit our proposed multivariate spatial copula model, we first need to retrieve stationary residuals that are uniformly distributed marginally. To this end, we preprocess our data site by site separately, first removing their seasonal pattern and then using the rank-based empirical distribution function to standardize the data to the uniform scale} (details are given in the Supplementary Material. 
For 2020, bivariate scatterplots of residuals obtained at each station are displayed in Figure \ref{fig:u_resi2020}. Even though the sample size is rather small, we can see that most pairs are moderately dependent and mildly asymmetric, with usually slightly stronger dependence in the upper tail (i.e., high daily temperature maxima with high daily minima). This suggests that our proposed model {might be} appropriate for capturing these dependence features.
\subsection{Model selection and diagnostics}
We fit the bivariate spatial model specified in \S\ref{subsec:illustration} to the data for the year 2020. {For the inference, we follow the Bayesian procedure presented in \S\ref{sec:inference}, but because the sample size is rather small (here, $n=271$), we adopt the two-step approach outlined in the simulation study in \S\ref{sec:simulation}.} Specifically, we fix $\alpha_1$, $\alpha_2\in\{0.5, 2, 5\}$ and $\delta^U, \delta^L\in \{0.3, 0.6, 0.9\}$, and then apply our customized MALA-based MCMC algorithm of \S\ref{sec:inference} to estimate the remaining parameters, for each $\{\alpha_1,\alpha_2,\delta^U,\delta^L\}$ parameter configuration. This means that in total, we have 81 different {candidate} model specifications and we {then} compare the fitted models based on the estimated $\chi$ metric. {Among the 81 candidate models, our} ``best" model{, which provides the best fit in terms of bivariate extremal dependence structures has $\alpha_1 = 2, \alpha_2 = 2, \gamma_1 = 0.39, \gamma_2 = 0.60, \delta^U = 0.9, \delta^L = 0.9, \lambda_1 = 888.8, \lambda_2 = 868.6$, and $\rho_{12} = 0.06$, which suggests {that the most plausible} extremal dependence {structures for the upper and the lower tails of the bivariate TMAX/TMIN spatial process might be respectively of the following types: U: AI, AD--AI and L: AD, AI--AI, using the notation introduced previously. In other words, while TMAX appears to be (spatially) asymptotically independent in the upper tail but asymptotically dependent in the lower tail, the opposite is true for TMIN, thus inducing asymptotic independence in both tails for the cross-dependence between TMAX and TMIN.} 

Then, we compare the model performance at finite threshold levels and different distances by plotting the extremal dependence measures ${\chi}_{11}(h, u)$, ${\chi}_{22}(h, u)$, and ${\chi}_{12}(h, u)$, estimated both empirically and based on our best fitted model, as a function of distance $h$ and threshold $u$. {Figure \ref{fig:app_U_09} shows the model performance {for} the upper tail, as a function of distance $h$ for a fixed threshold $u=0.9$ (top), and as a function of the threshold $u$ for a fixed distance $h=28.39$ (bottom), i.e., for two close-by sites. The results for medium and large distances, i.e., $h = 232.73$ and $h = 432.79$, are reported in the Supplementary Material. Overall, the plots show that the within- and cross-extremal dependence structures are captured accurately at all threshold levels and spatial distances, and that the model-based extremal dependence estimates have considerably lower uncertainty than their empirical counterparts.} Furthermore, our results indicate that {the} spatial dependence in each individual field is quite strong, but the cross-dependence structure is rather weak (but not inexistent).

\begin{figure}[t!]
	\centering
	\includegraphics[width=0.95\linewidth]{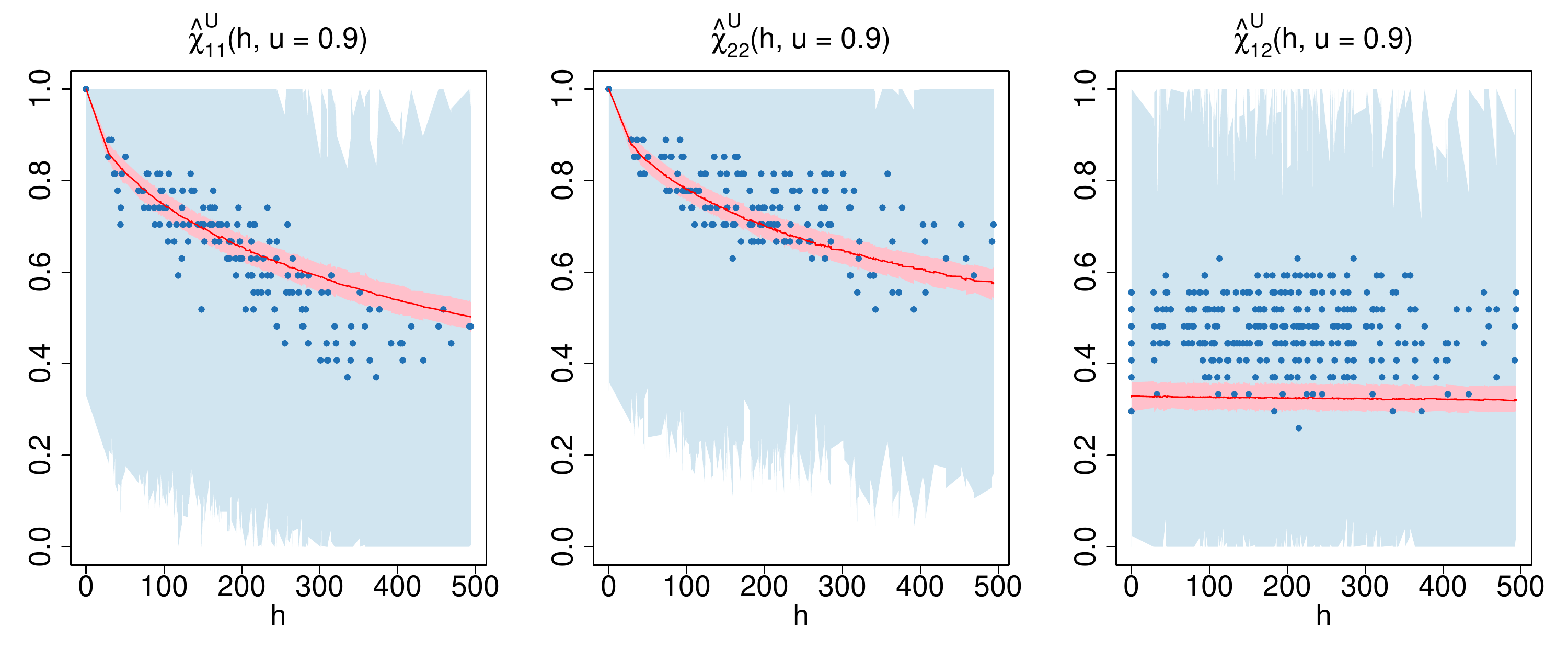}
	\includegraphics[width=0.95\linewidth]{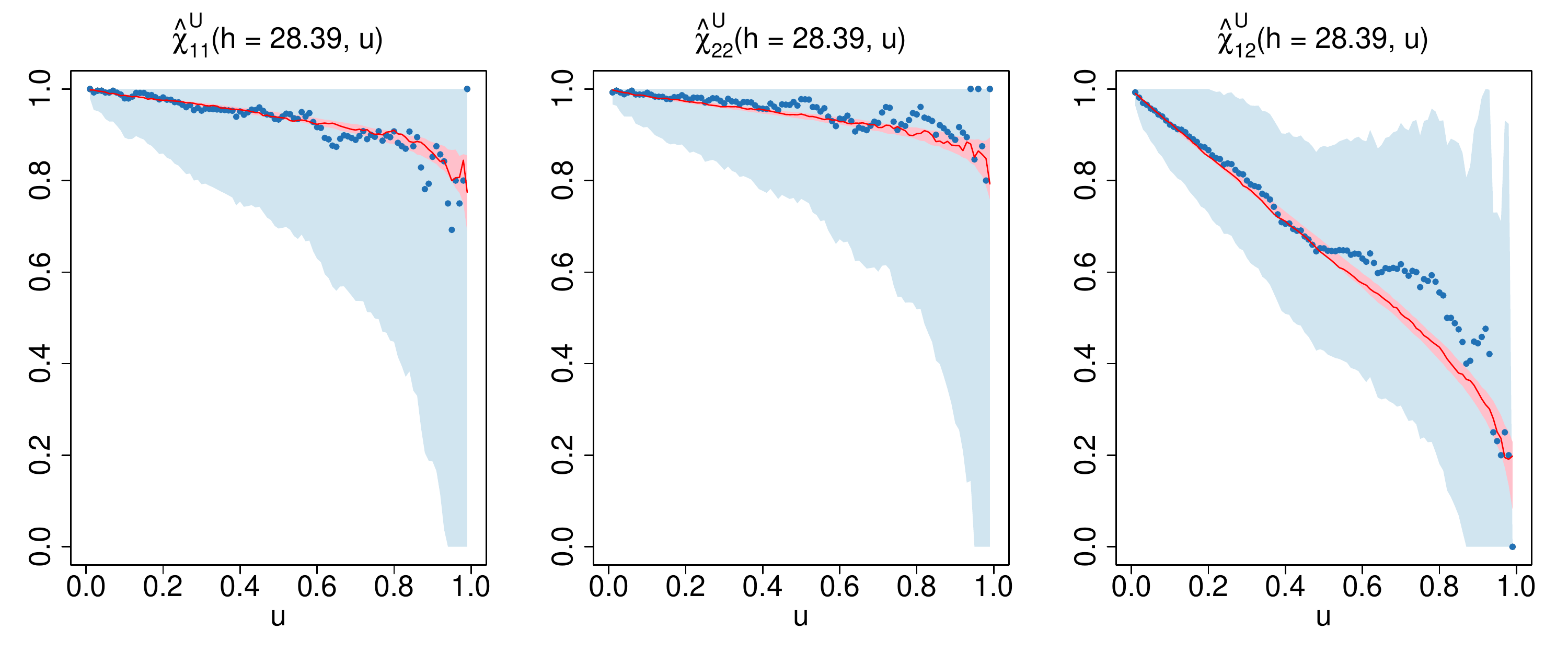}
	\caption{{Empirical estimates (blue dots), and model estimates (red lines) of ${\chi}_{11}^U(h, u)$, ${\chi}_{22}^U(h, u)$, ${\chi}_{12}^U(h, u)$ (left to right) as a function of distance $h$ for fixed $u=0.9$ (top), and as a function of threshold $u$ for fixed $h=28.39$ (bottom). The envelopes (blue shadows) are pointwise 95\% confidence intervals for the empirical estimates. The envelopes (pink shadows) are 95\% credible intervals sampled from the post-burn-in MCMC chains.}}\label{fig:app_U_09}
\end{figure}



Using the estimated model parameters, we can then simulate synthetic data at a finer resolution over the study region. Figure \ref{fig:post} shows one simulation result {from the fitted model,} displayed on the uniform scale. This simulation illustrates again that both TMAX and TMIN processes are large-scale phenomena, {with TMIN being more strongly dependent (spatially) than TMAX in the upper tail,} and that they are also moderately cross-dependent, as peaks and troughs appear to be co-located.

\begin{figure}[t!]
	\centering
	\includegraphics[width=0.85\linewidth]{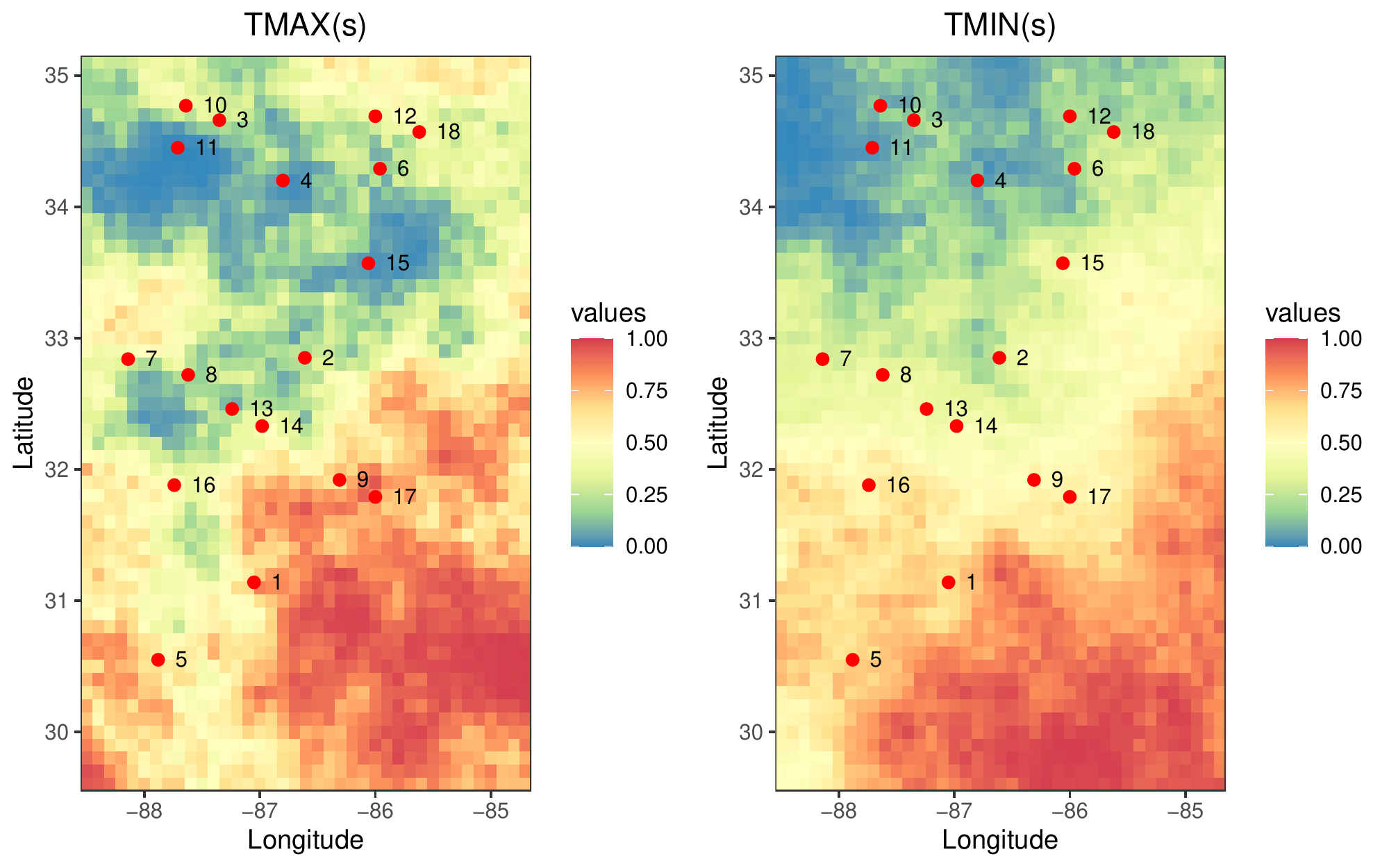}
	\caption{{Simulation from the fitted model, displayed on the uniform scale, on a fine-resolution grid covering the study region.}}
	\label{fig:post}
\end{figure}

                                   \section{Discussion}\label{sec:conclusion}
In this {paper}, we have proposed a novel multi-factor copula model for multivariate spatial data, with appealing tail dependence properties. The model is, {by design}, able to distinctly capture the different combinations of marginal and cross-extremal dependence types within and across different spatial random fields, with a smooth transition between tail dependence classes in the interior of the parameter space. {To the best of our knowledge, this is the first multivariate spatial model of this kind, contrasting} with all other multivariate spatial models from the published literature{, which either lack tail flexibility or have restrictive symmetries. Thanks to its flexibility and ability to handle tail asymmetry, our proposed model may be used in a wide range of real-world applications.} Although we {mainly focused in our study} on accurately capturing the joint lower and upper tails, we also stress that {the proposed model was fitted to the entire dataset (i.e., not only to one tail), and that its flexibility allows one to accurately capture the full distribution (bulk and tails) of multivariate spatial data.} Furthermore, inference can be performed using a Bayesian approach, which exploits the natural hierarchical representation of our model described by a simple Gaussian copula at the data level, and multiple exponential random factors at the latent level. We adopted a MALA-type algorithm with adaptive block proposals to boost its efficiency and enhance {the} convergence {of Markov chains to their} stationary distribution. In case {of high-dimensional data, computations of the required gradients} can be run in parallel to accelerate the inference procedure. {To further speed-up computations,} \cite{aicher2019stochastic} proposed the Stochastic gradient Langevin dynamics (SGLD) MCMC algorithm for models with latent factors, which we plan to explore in future research. 

{Finally, in our application, we investigated the spatial joint tail dependence structure of TMAX and TMIN in Alabama, US, and interestingly, our model revealed different asymptotic dependence types for each process and each tail. Although this finding is subject to uncertainty, it has important implications on environmental and heat-related health concerns, and it would be interesting to confirm our study in future research using a larger dataset.}

\section*{Acknowledgments}
This publication is based upon work supported by the King Abdullah University of Science and Technology (KAUST) Office of Sponsored Research (OSR) under Awards No. OSR-CRG2017-3434 and No. OSR-CRG2020-4394.


\baselineskip=12pt
\bibliographystyle{CUP}
\bibliography{Biblio}

\newpage
\appendix
\newpage
\section{{Basic model properties}}
\subsection{Marginal distribution functions}\label{appendix:Fxi}
Recall the model construction
\begin{align}\notag
X_i(\bs s) = \alpha_i\big\{ \gamma_i[\delta^UR_0^U+(1-\delta^U)R_i^U]-(1-\gamma_i)[\delta^LR_0^L+(1-\delta^L)R_i^L]\big\}+W_i(\bs s).
\end{align}
{Let}  
\begin{align}
\notag T_i  :&= \alpha_i\big\{ \gamma_i[\delta^UR_0^U+(1-\delta^U)R_i^U]-(1-\gamma_i)[\delta^LR_0^L+(1-\delta^L)R_i^L]\big\} \\
\notag & = \alpha_i \gamma_i \delta^UR_0^U + \alpha_i \gamma_i(1-\delta^U)R_i^U - \alpha_i(1-\gamma_i)\delta^LR_0^L - \alpha_i(1-\gamma_i)(1-\delta^L)R_i^L \\
\notag :&= \beta_{i0}^U R_0^U + \beta_{i}^U R_i^U - \beta_{i0}^L R_0^L -  \beta_{i}^L R_i^L,
\end{align}
where $\beta_{i0}^U = \alpha_i \gamma_i \delta^U, \beta_{i}^U = \alpha_i \gamma_i(1-\delta^U), \beta_{i0}^L = \alpha_i(1-\gamma_i)\delta^L$,  and $\beta_{i}^L = \alpha_i(1-\gamma_i)(1-\delta^L).$ Then the model can be written as 
$$X_i(\bs s) =  T_i + W_i(\bs s),$$
and it is straightforward to derive that the marginal distribution function of $X_{i}$ can be expressed as
\begin{equation}\label{eq:Fxi}
F^X_{i}(x)=\begin{cases}
\dfrac{1/2\exp(x)}{(\beta_{i0}^U+1)(\beta_{i0}^L-1)(\beta_{i}^U+1)(\beta_{i}^L-1)}
+\dfrac{(\beta_{i0}^L)^5\exp(x/\beta_{i0}^L)}{ (\beta_{i0}^L+1)(\beta_{i0}^L-1)(\beta_{i0}^L+\beta_{i0}^U)(\beta_{i0}^L+\beta_{i}^U)(\beta_{i0}^L-\beta_{i}^L)}\\
\\
\qquad+
\dfrac{(\beta_{i}^L)^5\exp(x/\beta_{i}^L)}{ (\beta_{i}^L+1)(\beta_{i}^L-1)(\beta_{i}^L+\beta_{i}^U)(\beta_{i}^L+\beta_{i0}^U)(\beta_{i}^L-\beta_{i0}^L)},
\qquad \text{for }\beta_{i0}^L\neq1, \beta_{i}^L\neq1, \beta_{i0}^L\neq\beta_{i}^L,
& w< 0,\\ 
\\
1-\dfrac{1/2\exp(-x)}{ (\beta_{i0}^U-1)(\beta_{i0}^L+1)(\beta_{i}^U-1)(\beta_{i}^L+1)}
-\dfrac{(\beta_{i0}^U)^5\exp(-x/\beta_{i0}^U)}{ (\beta_{i0}^U+1)(\beta_{i0}^U-1)(\beta_{i0}^L+\beta_{i0}^U)(\beta_{i0}^U+\beta_{i}^L)(\beta_{i0}^U-\beta_{i}^U)}\\
\\
\qquad-
\dfrac{(\beta_{i}^U)^5\exp(-x/\beta_{i}^U)}{ (\beta_{i}^U+1)(\beta_{i}^U-1)(\beta_{i}^L+\beta_{i}^U)(\beta_{i}^U+\beta_{i0}^L)(\beta_{i}^U-\beta_{i0}^U)},
\qquad \text{for }\beta_{i0}^U\neq1, \beta_{i}^U\neq1, \beta_{i0}^U\neq\beta_{i}^U,
& w\geq 0.\\
\end{cases}
\end{equation}
All other cases can be derived as limits.
\subsection{The covariance structure of the bivariate $(W_1(\bs s),W_2(\bs s))^\top$ process}\label{appendix:sigma}
Recall the model construction and notation in Section~\ref{sec:modelconstruction}. The components of the mean vector are
	\begin{align}
		\notag \E(\bs W_{1}' )=&\E(\bs {W}_{1}^*) = \bs 0_d,\\
		\notag \E(\bs W_{2}' )=&\E(\rho_{12}{\bs W}_{1}^*
		+\sqrt{1-\rho_{12}^2}\bs {W}_{2}^*)=\rho_{12}\E(\bs {W}_{1}^*)+\sqrt{1-\rho_{12}^2}\E(\bs {W}_{2}^*)= \bs 0_d,
	\end{align}
	and the covariance matrix $\Sigma_{2d\times 2d}$ can be constructed as
	\[
		\begin{bmatrix}
			\Sigma^{'}_{1} &\Sigma^{'}_{12}\\
			\Sigma^{'}_{12} &\Sigma^{'}_{2}
		\end{bmatrix},
	\]
	where the elements of the block covariance matrix are
	\begin{align}
		\notag  \Sigma^{'}_{1,jj}= &\;\var(W_{1j}')=\var({W}_{1j}^*) = 1,\\
		\notag  \Sigma^{'}_{1,jk}= &\;\cov(W_{1j}',W_{1k}') ={\Sigma}_{1,jk}^* =c_1(h_{jk}),\\
		\notag  \Sigma^{'}_{2,jj}= &\;\var(W_{2j}')=\var(\rho_{12}{W}_{1j}^*
		+\sqrt{1-\rho_{12}^2}{W}_{2j}^*) = \rho_{12}^2\var({W}_{1j}^*)
		+(1-\rho_{12}^2)\var({W}_{2j}^*) = 1,\\
		\notag  \Sigma^{'}_{2,jk} =&\;\cov(W_{2j}', W_{2k}')= \cov(\rho_{12}{W}_{1j}^*
		+\sqrt{1-\rho_{12}^2}{W}_{2j}^*,\rho_{12}{W}_{1k}^*
		+\sqrt{1-\rho_{12}^2}{W}_{2k}^*)\\
		\notag= &\;
		\rho_{12}^2\cov({W}_{1j}^*,{W}_{1k}^*)
		+(1-\rho_{12}^2)\cov({W}_{2j}^*,{W}_{2k}^*)\\
		\notag=&\;\rho_{12}^2c_1(h_{jk})+(1-\rho_{12}^2)c_2(h_{jk}),\\
		\notag  \Sigma^{'}_{12,jj}= &\;\cov(W_{1j}', W_{2j}')=\cov({W}_{1j}^*,
		\rho_{12}{W}_{1j}^*
		+\sqrt{1-\rho_{12}^2}{W}_{2j}^*
		) = \rho_{12}\var({W}_{1j}^*)= \rho_{12},\\
		\notag \Sigma^{'}_{12,jk} = &\;\cov(W_{1j}', W_{2k}')=\cov({W}_{1j}^*,
		\rho_{12}{W}_{1k}^*
		+\sqrt{1-\rho_{12}^2}{W}_{2k}^*
		) = \rho_{12}\cov({W}_{1j}^*,{W}_{1k}^*)=\rho_{12}c_1(h_{jk}),
	\end{align}
and $j,k = 1,\dots,d.$

\section{Proof of tail dependence properties}\label{appendix:proof}

We start with the formal proof of the spatial tail dependence properties of each individual process $X_i(\bs s)$. Following Appendix \S\ref{appendix:Fxi}, the model can be written as $$X_i(\bs s) =  T_i + W_i(\bs s),$$
and the cumulative distribution function of $T_i$ can be expressed in closed form, namely
\begin{align}\label{eq:Fti}
F_{T_i}(t) 
& =
\notag 
\begin{cases}
\dfrac{(\beta_{i0}^L)^3 \exp(t/\beta_{i0}^L)}
{(\beta_{i}^U+\beta_{i0}^L)(\beta_{i0}^L+\beta_{i0}^U)(\beta_{i0}^L-\beta_{i}^L)}
-\dfrac{(\beta_{i}^L)^3\exp(t/\beta_{i}^L)}{(\beta_{i}^U+\beta_{i}^L)(\beta_{i0}^L-\beta_{i}^L)(\beta_{i}^L+\beta_{i0}^U)},
\quad & t< 0,\\ 
\\
1-
\dfrac{(\beta_{i0}^U)^3\exp(-t/\beta_{i0}^U)}
{(\beta_{i0}^L+\beta_{i0}^U)(\beta_{i0}^U-\beta_{i}^U)(\beta_{i}^L+\beta_{i0}^U)}
-\dfrac{(\beta_{i}^U)^3\exp(-t/\beta_{i}^U)}{(\beta_{i}^U+\beta_{i0}^L)(\beta_{i0}^U+\beta_{i}^U)(\beta_{i0}^U-\beta_{i}^U)},
\quad & t\geq 0.\\ 
\end{cases}
\\
\notag  &\\
:& =
\notag 
\begin{cases}
A_{i0}\exp(t/\beta_{i0}^L) - A_i \exp(t/\beta_{i}^L),\quad & t< 0,\\ 
\\
1-B_{i0}\exp(-t/\beta_{i0}^U) - B_i \exp(-t/\beta_{i}^U),\quad & t\geq 0.\\ 
\end{cases}
\end{align}
where  $$A_{i0} =\dfrac{(\beta_{i0}^L)^3}{(\beta_{i}^U+\beta_{i0}^L)(\beta_{i0}^L+\beta_{i0}^U)(\beta_{i0}^L-\beta_{i}^L)}, \quad A_i = \dfrac{(\beta_{i}^L)^3}{(\beta_{i}^U+\beta_{i}^L)(\beta_{i0}^L-\beta_{i}^L)(\beta_{i}^L+\beta_{i0}^U)},$$
$$B_{i0} = \dfrac{(\beta_{i0}^U)^3}
{(\beta_{i0}^L+\beta_{i0}^U)(\beta_{i0}^U-\beta_{i}^U)(\beta_{i}^L+\beta_{i0}^U)}, \quad B_i = \dfrac{(\beta_{i}^U)^3}{(\beta_{i}^U+\beta_{i0}^L)(\beta_{i0}^U+\beta_{i}^U)(\beta_{i0}^U-\beta_{i}^U)}.$$
By taking the exponential on both components of $X_i(\bs s)$, we obtain $\tilde{X_i}(\bs s)$ = $\tilde{T_i}$ $\tilde{W_i}(\bs s)$, where $\tilde{T_i} = \exp(T_i)$ and $\tilde{W_i}(\bs s) = \exp(W_i(\bs s))$, for $i = 1,\dots,p$. Then,  for $t>1$, we have 
$$
\Pr(\tilde{T_i} > t) = \Pr(T_i>\log t) = B_{i0} t^{-1/\beta_{i0}^U} + B_i t^{-1/\beta_{i}^U}
\stackrel{t\rightarrow \infty}{\sim}\quad
\notag 
\begin{cases}
B_{i0} t^{-1/\beta_{i0}^U}, & \text{if } \beta_{i0}^U>\beta_{i}^U,\\
B_i t^{-1/\beta_{i}^U}, & \text{if } \beta_{i0}^U<\beta_{i}^U.
\end{cases}
$$
Therefore, the value $\max(\beta_{i0}^U, \beta_{i}^U)$ determines the decay rate of $\Pr(\tilde{T_i} > t)$, and $\tilde{T_i}$ is regularly-varying at infinity with index $-\alpha_{\tilde{T}_i}$, with $\alpha_{\tilde{T}_i}=1/\max(\beta_{i0}^U,\beta_{i}^U)$.

Now, for $w>1$, $\Pr(\tilde{W_i} > w) = \Pr(W_i>\log w) = 1/(2w)$, which implies that $\tilde{W_i}$ is regularly-varying at infinity with index $-1$. Moreover, $\Pr(\tilde{W_i} > 0) = 1,$ for $i=1, \ldots, p.$ Let $\varepsilon>0$ and define $\tilde{\varepsilon} = \varepsilon/\alpha_{\tilde{T_i}}>0$,
\begin{align}
\E(\tilde{W_i}^{\alpha_{\tilde{T}_i}+\varepsilon}) 
\notag&= \int_0^{\infty} \Pr(\tilde{W}_i^{\alpha_{\tilde{T}_i}+\varepsilon}>w)\text{d}w 
= \int_0^{\infty} \Pr(\tilde{W}_i>w^{1/\{\alpha_{\tilde{T}_i}(1+\tilde{\varepsilon})\}})\text{d}w \\
\notag&= \int_0^{1} \Pr(\tilde{W}_i>w^{1/\{\alpha_{\tilde{T}_i}(1+\tilde{\varepsilon})\}})\text{d}w + \dfrac{1}{2}\int_1^{\infty} w^{-1/\{\alpha_{\tilde{T}_i}(1+\tilde{\varepsilon})\}}.
\end{align}
The first integral is bounded above by 1 and the second integral is finite if and only if $1/\{\alpha_{\tilde{T}_i}(1+\tilde{\varepsilon})\}>1$, i.e., $\alpha_{\tilde{T}_i}<1/(1+\tilde{\varepsilon})$. 

Then, by letting $\varepsilon\to0$, we conclude from Table 2 of \citet{engelke2019extremal} that when $\alpha_{\tilde{T}_i}<1$, or equivalently when $\max(\beta_{i0}^U,\beta_{i}^U)=\alpha_i\gamma_i\max(\delta^U,1-\delta^U)>1$, the vector $(X_{i}(\bs s_{j_1}), X_{i}(\bs s_{j_2}))^\top$ (which has the same copula structure as $(\tilde{X}_{i}(\bs s_{j_1}), \tilde{X}_{i}(\bs s_{j_2}))^\top$) is AD. On the other hand, when $\alpha_{\tilde{T}_i}>1$, because $(W_{i}(\bs s_{j_1}), W_{i}(\bs s_{j_2}))^\top$ is Gaussian with correlation (and thus has $\chi_U=0$ for correlations less than one, according to \citet{sibuya1960bivariate} and \citet{ledford1996statistics}), we deduce from Proposition 5 of \citet{engelke2019extremal} that the vector $(X_{i}(\bs s_{j_1}), X_{i}(\bs s_{j_2}))^\top$ has $\chi_U=0$, i.e., it is AI. The case $\alpha_{\tilde{T}_i}=1$ can be deduced by applying Proposition 6(3c) of \citet{engelke2019extremal}. Similar arguments hold for the lower tail.

For the cross-dependence structure, we argue by analogy. The vector $(X_{i_1}(\bs s_{j_1}), X_{i_2}(\bs s_{j_2}))^\top$ has the same dependence structure as the vector $(\tilde{X}_{i_1}(\bs s_{j_1}), \tilde{X}_{i_2}(\bs s_{j_2}))^\top$, for $i, i_1, i_2 = 1,\dots,p$ and $j_1, j_2 = 1,\dots,d$. Both components share the random variables $R_0^U$ and $R_0^L$, which are the only terms that can induce AD in the upper and lower tails, respectively, provided their respective coefficients are large enough. The term $R_0^U$ is the dominating term in the upper tail decay of both processes when we have $\beta_{i_1 0}^U>\max\{\beta_{i_1}^U, 1\}$ and $\beta_{i_2 0}^U>\max\{\beta_{i_2}^U, 1\}$ simultaneously, i.e., $\delta^U>1-\delta^U$ (i.e., $\delta^U>1/2$), $\alpha_{i_1}\gamma_{i_1}\delta^U>1$ and $\alpha_{i_2}\gamma_{i_2}\delta^U>1$. The result follows. A similar argument holds for the lower tail. To summarize, we display the marginal and cross tail dependence properties in Table \ref{tab:tailproperties}.

\end{document}